\documentclass[prx,twocolumn,showpacs,nofootinbib]{revtex4-1}
\usepackage{dsfont}
\usepackage{amsfonts,amsmath,amssymb}
\usepackage[english]{babel}
\usepackage{graphicx}
\usepackage{bm}
\usepackage[caption=false]{subfig}
\usepackage[dvipsnames]{xcolor}
\usepackage[hyperindex,breaklinks]{hyperref}
\usepackage{multirow}

\graphicspath{ {./images/} }
\def\cH{{\mathcal{H}}}

\def\j{{\sf{j}}}
\def\l{{\sf{l}}}

\def\bk{{\mathbf{k}}}
\def\bx{{\mathbf{x}}}
\def\bc{{\mathbf{c}}}
\def\bp{{\mathbf{p}}} 
\def\bP{{\mathbf{P}}} 
\def\bh{{\mathbf{h}}} 
\def\bq{{\mathbf{q}}} 
\def\br{{\mathbf{r}}}
\def\bs{{\mathbf{s}}}
\def\bd{{\mathbf{d}}}
\def\bR{{\mathbf{R}}}
\def\bB{{\mathbf{B}}}
\def\bxi{{\boldsymbol \xi}} 
\def\bkappa{{\boldsymbol \kappa}}
\def\tbkappa{{\tilde \bkappa}}
\def\bsigma{{\boldsymbol \sigma}}

\def\bQp{{ \mathbf {\hat Q}}_\perp}

\def\bsigma{{\boldsymbol \sigma}} 
\def\no{\nonumber} 
\def\k{{\kappa}}

\def\up{{\uparrow}}

\def\down{{\downarrow}}
\def\a{{\alpha}}

\def\g{{\gamma}}
\def\d{{\delta}}
\def\D{{\Delta}}
\def\sd{{\mathcal{D}}}

\def\cH{{\mathcal{H}}}

\def\dot{{\!\,\cdot\!\,}}

\newcommand{\ket}[1]{| #1 \rangle}
\newcommand{\bra}[1]{\langle #1 |}

\newcommand{\ignore}[1]{ }

\begin{document}
\preprint{Preprint number}

\title{Quantum Entangled-Probe Scattering Theory}

\author{Abu Ashik Md Irfan}
\email{airfan@iu.edu}
\affiliation{Department of Physics, Indiana University, Bloomington, IN 47408, USA}
\affiliation{Quantum Science and Engineering Center, Indiana University, Bloomington, IN 47408, USA}

\author{Patrick Blackstone} 
\email{pblackst@iu.edu}
\affiliation{Department of Physics, Indiana University, Bloomington, IN 47408, USA}
\affiliation{Quantum Science and Engineering Center, Indiana University, Bloomington, IN 47408, USA}

\author{Roger Pynn}
\email{rpynn@iu.edu}
\affiliation{Department of Physics, Indiana University, Bloomington, IN 47408, USA}
\affiliation{Quantum Science and Engineering Center, Indiana University, Bloomington, IN 47408, USA}

\author{Gerardo Ortiz}
\email{ortizg@iu.edu}
\affiliation{Department of Physics, Indiana University, Bloomington, IN 47408, USA}
\affiliation{Quantum Science and Engineering Center, Indiana University, Bloomington, IN 47408, USA}

\begin{abstract} 
\noindent 
We develop an entangled-probe scattering theory, including quantum detection, that extends the scope of standard scattering approaches. We argue that these probes may be revolutionary in studying entangled matter such as unconventional phases of strongly correlated systems. Our presentation focuses on a neutron beam probe that is mode-entangled in spin and path as is experimentally realized in~\cite{shen2019unveiling}, although similar ideas also apply to photon probes. We generalize the traditional van Hove theory \cite{vanhove1954} whereby the response is written as a properly-crafted combination of two-point correlation functions. Tuning the probe's entanglement length allows us to interrogate spatial scales of interest by analyzing interference patterns in the differential cross-section. Remarkably, for a spin dimer target we find that the typical Young-like interference pattern observed if the target state is un-entangled gets quantum erased when that state becomes maximally entangled.

\end{abstract} 

\date{\today}
\maketitle



\section{Introduction}

For more than a century, scattering techniques have been successfully employed to extract information about structural and dynamical properties of matter. Different types of probes (X-rays, electrons, neutrons, for example) reveal different (classical or quantum) characteristic properties of the target system depending on the nature of the probe-target interaction. So far, no quantum probe has exploited the characteristic trait of quantum mechanics: entanglement. Can one realize entangled-beams of particles? What information do entangled probes extract from the target? In this work we develop an entangled-probe scattering theory that addresses some of these issues at a fundamental level. \ignore{Application to magnetic scattering by a simple dimer shows that completely different scattering signatures emerge when both the probe and the dimer spins are entangled.}

\begin{figure}[t]
    \centering
    \includegraphics[height=0.65\columnwidth]{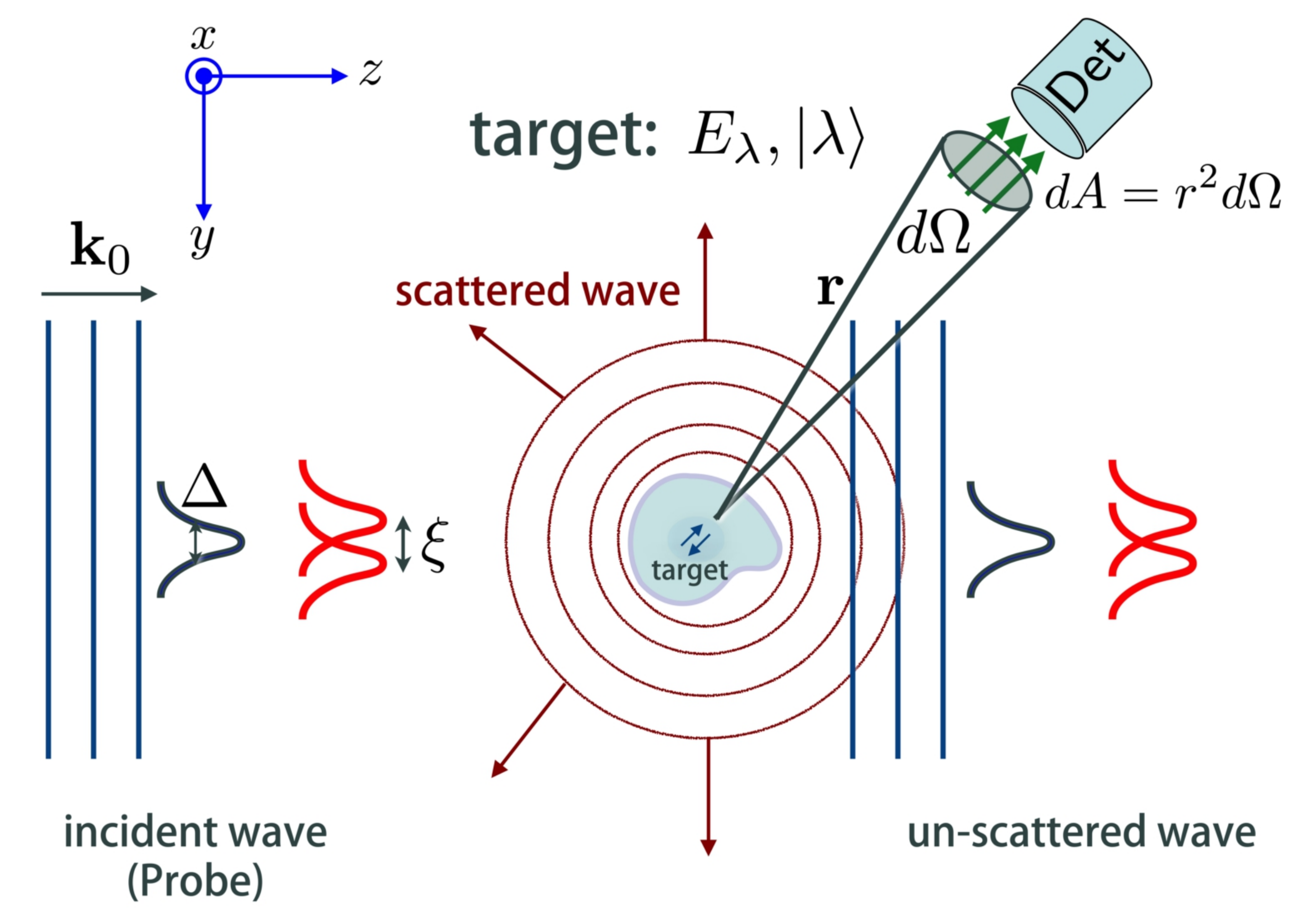}
    \caption{Scattering layout for the entangled probe of entanglement length $\xi$ compared to an un-entangled wave packet, of transverse width $\Delta$, and a plane wave, of momentum $\bk_0$. Scattered waves from the target, with energies  $E_\lambda$ and eigenstates $\ket{\lambda}$, are detected at distance $r$ and solid angle $d\Omega$.}
    \label{fig:entangledscattering}
\end{figure}

Recent work~\cite{shen2019unveiling,lu2019operator,kuhn2020unveiling} has demonstrated two types of {\it entanglers} capable of preparing a beam of neutrons in a state exhibiting mode entanglement in two (spin-path) or three (spin-path-energy) distinguishable subsystems. These probes can (and will) be used in scattering experiments to examine condensed matter systems in a way similar to standard neutron scattering~\cite{loveseybook}. Ideally, one would like to develop quantum measurements that identify/detect the entanglement present in the target matter. Thus, extension of the standard textbook theory of scattering \cite{Sakuraibook, newtonbook,joachain} to include entanglement of the probe (or projectile) is necessary. Typically, projectiles are counted by detectors arranged spatially (see Fig.~\ref{fig:entangledscattering}). The nature of those detectors may vary depending on the property of the projectile one is trying to unveil, and the counting rate as a function of scattering angle from the direction of incidence defines the differential cross section (DCS). 

Mode entanglement in the state of a single particle refers to the existence of non-local correlations between its different distinguishable subsystems (path, spin, energy, etc.);  alternatively, if non-local correlations between multiple particles are present one speaks of particle entanglement \cite{barnum-2003, barnum-2004}. The latter is realized, for instance, in beams of entangled photon pairs \cite{Schotland2016}. As we will see, these two different forms of scattering probe entanglement differ in the type of information they extract from the target. While mode-entangled scattering uncovers distinctively crafted two-point correlations of the sample being probed, matrix elements in multiparticle scattering include two-body interaction operators for each particle, thus contributing to a multi-point correlation function. Importantly, by tuning the probe's entanglement one can in principle vary the correlation function and unveil entanglement in the target system.

A main result of this paper is the formulation of a scattering theory for a mode-entangled probe. When applied to a magnetic interaction potential, such formulation represents an extension of van Hove's theory.  To gain understanding of the kind of information one can extract using this kind of probe, we illustrate our findings in the case of a spin-dimer. Depending on the relation between the various length scales involved in the scattering process, one obtains radically different responses allowing detection of entanglement in the target material.  For instance, the typical Young-like interference pattern observed when the target state is un-entangled gets {\it quantum erased} when the target state is maximally entangled. In addition, to obtain ancillary information on the nature of entanglement in the target we propose to alter the way one detects the outgoing scattered probe, as happens when one measures spin polarization using a spin-echo device in neutron scattering.

The paper is organized as follows. Section~\ref{sec:epst} starts with the formal introduction of a multimode-entangled probe and then sets the scattering framework by generalization of the standard $T$-matrix formalism (details of the derivation can be found in Appendix \ref{appendix:born-calculation}). Section~\ref{sec:magnetic scattering} discusses the particular case of magnetic interaction between the target and a neutron probe and derives the extension of van Hove's theory. To identify the kind of information the entangled probe can extract from a target state, this magnetic scattering is further specialized in Section~\ref{sec:dimer} using the example of a spin-dimer target. In this section, we illustrate how the DCS changes under different experimental conditions and discuss the prospects of detecting the varied outcomes. Finally, in Section~\ref{sec:Outlook} we summarize main results and implications of this work. An Appendix \ref{appendix:multiparticle} briefly discusses the general scattering theory for the multiparticle-entangled fermion-probe case.

\section{Entangled-Probe Scattering Theory}\label{sec:epst}

For simplicity, we focus on mode entanglement and consider a coherent beam with two distinguishable subsystems: path and spin. In distinguishing these subsystems, the full Hilbert space of our probe state is constructed as ${\cal H}_{\sf probe}={\cal H}_{\sf path}\otimes {\cal H}_{\sf spin}$. The two pathways of the probe are indistinguishable alternatives, and given a separation in paths $\xi$ (Fig.~\ref{fig:entangledscattering}), a wave packet description must be employed. 

\begin{figure}[t]
    \centering 
    \includegraphics[width=\columnwidth]{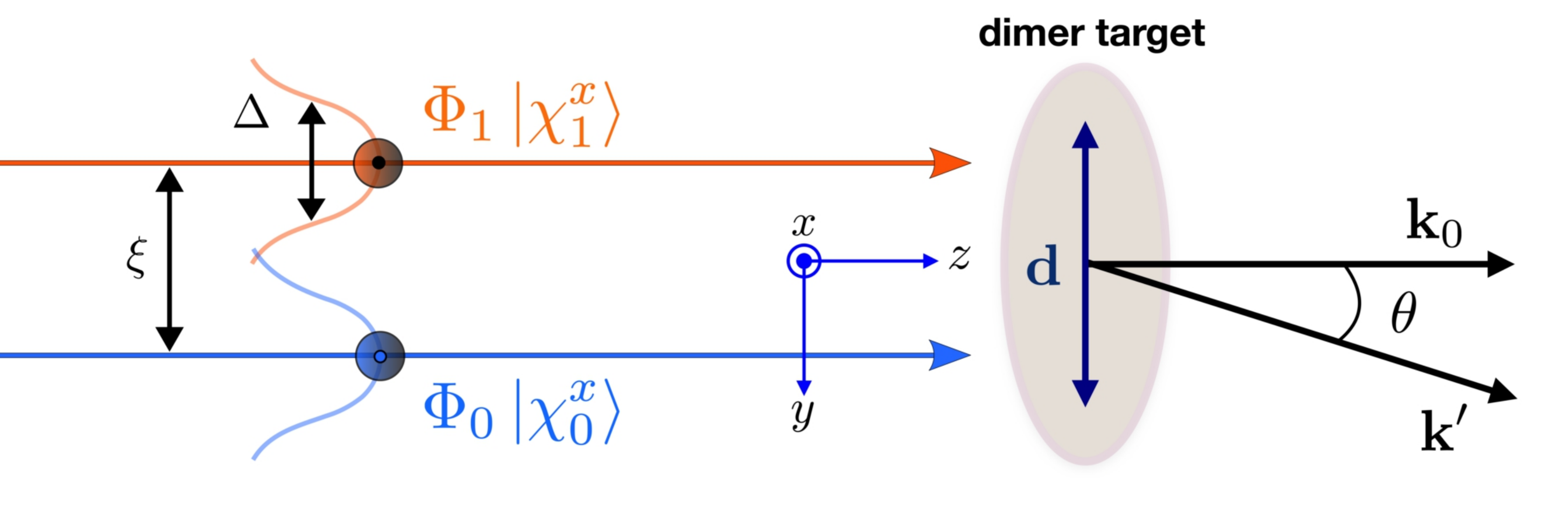}
    \caption{Schematic of probe wave packets of mean momentum $\bk_0$, illustrating that the spin and path subsystems are entangled. This figure also shows the scattering to momentum $\bk'$ by a particular orientation of the dimer vector, $\bd = | \bd | \hat{y}$, which is investigated in Section~\ref{sec:dimer}.}
    \label{fig:dimer orientation specific}
\end{figure}

We define the simplest initial, $t_0<0$, single-particle entangled-probe state to be 
\begin{eqnarray}
        \Phi_\text{in}(\br,t_0)
    &=& \frac{\Phi_0(\br,t_0) \ket{\chi^\alpha_0} + \Phi_1(\br,t_0) \ket{\chi^\alpha_1}}{\sqrt{2}}, 
\end{eqnarray}
(see Fig.~\ref{fig:dimer orientation specific}). Equivalently, in plane wave components \cite{1Note}
\begin{eqnarray}
        \Phi_\text{in}(\br,t_0)
    &=& \frac{1}{L^\frac{3}{2}}\sum_\bk  \tilde g(\bk) e^{i \bk \cdot \br - i \omega(k) t_0} \ \ket{\chi_{\bk\cdot\bxi}}, 
    \no
    \\
    \mbox{ with } \quad  \ket{\chi_{\bk\cdot\bxi}}&=& 
    \frac{e^{- \frac{i}{2} \Theta_\bk} \ket{\chi^\alpha_0} + e^{\frac{i}{2} \Theta_\bk} \ket{\chi^\alpha_1}}{\sqrt{2}},
    \no
\end{eqnarray}
where $\hbar \omega(k) = E_\bk = \hbar^2 k^2 / 2 m$, $m$ is the mass of the probe, 
$\Theta_\bk={\bk\!\cdot\!\bxi}+2\phi$, $\phi$ is a phase introduced by the experimental apparatus (i.e., the entangler~\cite{shen2019unveiling}), and the quantization 
box of the momentum $\hat \bp$ states $\ket{\bk}$ is taken to be of linear size $L$. 
We choose normalizations such that 
\begin{eqnarray}
    \langle \br \ket{\bk}=e^{i \bk \cdot \br}/L^\frac{3}{2}, \quad \langle \bk'\ket{\bk} = \delta_{\bk'\bk}, \quad \langle \br' \ket{\br} = \delta(\br' - \br). \no
\end{eqnarray}
The entangled wave packet is characterized by the distribution $\tilde g(\bk)$ with mean momentum $\bk_0$, 
transverse spatial width $\Delta$, and energy 
$\bra{\Phi_\text{in}}\hat H_{\sf p}\ket{\Phi_\text{in}}=E_{\sf p}$, with 
$\hat H_{\sf p}=\frac{\hat \bp^2}{2 m}$. The probe $ \Phi_\text{in}$ becomes unentangled with respect to the 
${\cal H}_{\sf probe}$ decomposition if $\xi = 0$.  The orthogonal spin-1/2 states $\ket{\chi^\alpha_{\nu}}$, $\nu=0,1$,  are 
defined along a particular spin-quantization axis $\alpha=x,y,z$,
\begin{eqnarray}
    \sigma^\alpha \ket{\chi^\alpha_{\nu}} = (-1)^\nu \ket{\chi^\alpha_{\nu}}, \qquad \hat{\bsigma} = (\sigma^x, \sigma^y, \sigma^z), \no
\end{eqnarray}
with $\sigma^\alpha$ representing Pauli matrices.
Starting from states quantized along $\alpha$, the effective spinor associated with the $\bk$ component of $\Phi_{\sf in}$ will be fully aligned along a particular direction given by
\begin{eqnarray}
    (\hat{\bsigma} \cdot \hat{\chi}_\alpha) \ket{\chi_{\bk \cdot \bxi}} = \ket{ \chi_{\bk \cdot \bxi} } \no
\end{eqnarray}
where the $(\bk \cdot \bxi)$-dependent axes $\hat{\chi}_\alpha$ are found to be
\begin{eqnarray}
    \hat{\chi}_x &=& ( 0, -\sin \Theta_\bk, \cos \Theta_\bk ) , \no \\
    \hat{\chi}_y &=& ( \sin \Theta_\bk, 0, \cos \Theta_\bk ) , \no \\
    \hat{\chi}_z &=& ( \cos \Theta_\bk, \sin \Theta_\bk, 0 ). \no
\end{eqnarray}

Since we are interested in both  elastic {\it and} inelastic scattering we must include 
dynamical properties of the target Hamiltonian $\hat H_{\sf t}$ of spectral 
representation $\hat H_{\sf t} \ket{\lambda} = E_\lambda \ket{\lambda}$. 
Then, the total Hamiltonian of the probe-target system is $\hat H=\hat H_0+\hat V$, where 
$\hat H_0=\hat H_{\sf p}+\hat H_{\sf t}$, with
$\hat V$ representing their interaction potential. 
The Hilbert space of the probe-target system is 
$\mathcal{H}_{\sf probe} \otimes \mathcal{H}_{\sf target}$,
and has basis elements denoted by $\ket{\bk \chi \lambda} \equiv \ket{\bk} \otimes \ket{\chi} \otimes \ket{\lambda}$.
We assume that the probe-target initial state is the mixed state $\hat \rho_{\sf in}=
\ket{\Phi_{\sf in}}\bra{\Phi_{\sf in}}\otimes \hat \rho_{\sf t}$, with $\hat \rho_{\sf t}=\sum_\lambda 
p_\lambda \ket{\lambda}\bra{\lambda}$, where $p_\lambda$ is a Boltzmann weight if 
the target is in thermodynamic equilibrium at $t=t_0$. 

We next extend  the $T$-matrix formalism to include entanglement in the probe. 
In the interaction picture the propagator obeys 
$    \hat U_I(t,t_0) = \mathds{1} - \frac{i}{\hbar} \int_{t_0}^t dt_1 \hat V_I(t_1) \hat U_I(t_1,t_0)
$, 
with $\hat U_I(t_0,t_0)=\mathds{1}$ and $\hat V_I(t)=e^{i\hat H_0t/\hbar} \hat V 
e^{-i\hat H_0 t/\hbar}$.  
The $T$ matrix, describing the transition 
from the state $\ket{\psi_\bk}=\ket{\bk \, \chi_{\bk\cdot \bxi}\, \lambda}$ to 
the basis state $\ket{\psi'}=\ket{\bk'\chi' \lambda'}$, is defined by
\begin{eqnarray}
    \bra{\psi'}U_I(t,t_0)\ket{\psi_\bk}&=&\delta_{\bk \bk'} \delta_{\lambda \lambda'} \langle\chi'\ket{\chi_{\bk\cdot \bxi}} 
    \no\\
    && 
    - \frac{i}{\hbar} \tilde T_{\psi'\psi_\bk}  \int_{t_0}^t dt_1 e^{i \omega(\psi', \psi_\bk)t_1 + \epsilon t_1} \no ,
\end{eqnarray}
in which we notate $\tilde T_{\psi'\psi_\bk}=\langle \psi' | \hat{T} | \psi_\bk \rangle$,
$\hbar \omega(\psi',\psi_\bk) \equiv E_{\psi'}-E_{\psi_\bk}$, 
$E_\psi=\bra{\psi}\hat H_0\ket{\psi}$, and $\epsilon>0$ a regulator. In the equation above  we have introduced the matrix element
per quantization volume $L^3$; eventually one has to perform the $L \to \infty$ limit. 
 The un-scattered state is decomposed in terms of the $\ket{\psi_\bk}$ states as
\begin{eqnarray} 
    \ket{\psi} = \sum_\bk \tilde g(\bk) \, \ket{\psi_\bk }=\ket{\Phi_{\sf in}}\otimes \ket{\lambda}.
\end{eqnarray} 
Note that this state does not contain $t$ dependence explicitly, as we are working in the interaction picture. 
Then, 
\begin{multline}
	\bra{\psi'} U_I(t,-\infty) \ket{\psi} =  \delta_{\lambda \lambda'} \tilde g(\bk')  \langle\chi'\ket{\chi_{\bk'\cdot \bxi}}
	\\
	- \frac{i}{\hbar}\sum_\bk  \tilde T_{\psi'\psi_\bk} \tilde g(\bk) \frac{e^{i \omega(\psi',\psi_\bk)t + \epsilon t}}{i \omega(\psi',\psi_\bk) + \epsilon} .
	\nonumber
\end{multline} 
Although the wave packet is not monoenergetic by design, the spread of $\tilde{g}(\bk)$ about its central value $\bk_0$ ($=k_0 \hat{z}$ in the geometry of Fig.~\ref{fig:entangledscattering}) is supposed to be small. As such, when measuring scattering away from the  propagation axis of the wave packet, $\bk'$ is far from $\bk_0$ and so $\tilde g(\bk')\approx 0$~\cite{Sakuraibook,newtonbook}. The forward scattering term, $\delta_{\lambda \lambda'} \tilde{g}(\bk') \langle\chi' \ket{\chi_{\bk' \cdot \bxi} }$, is thus omitted. Using this approximation and the density of states $\rho(E_{\bk'})= \frac{m k'}{\hbar^2} \left( \frac{L}{2\pi} \right)^3 d\Omega_{\bk'}$ with $d\Omega_{\bk'}$ the solid angle in the direction of $\bk'$, the 
probability of transition per final-state energy is shown to be (see Appendix \ref{appendix:born-calculation}) 
\begin{eqnarray}        
	&& \rho(E_{\bk'}) \lim_{t \to \infty} \int_{-\infty}^t dt' W_{\psi \to \psi'}(t') =  \label{eq:transitionprobability}   \\
	&& \hspace{0.5cm} \frac{2 \pi^2}{ \hbar^2} \rho(E_{\bk'}) \! \sum_{\bk_1, \bk_2} \!\! \tilde g^*(\bk_1) \tilde g(\bk_2)  \tilde T^*_{\psi' \psi_{\bk_1}} \! \tilde T_{\psi' \psi_{\bk_2}} \no \\
	&& \hspace{0.8cm} \times \delta\big(\omega(k_1)-\omega(k_2)\big) \big[\delta\big(\omega(\psi',\psi_{\bk_1})\big)+\delta\big(\omega(\psi',\psi_{\bk_2})\big)\big] \no ,
\end{eqnarray} 
wherein 
\begin{eqnarray}
	W_{\psi \to \psi'}(t) &\equiv& \lim_{\epsilon \to 0^+} \frac{d}{dt} |\bra{\psi'} U_I(t, -\infty) \ket{\psi} |^2 \no
\end{eqnarray}
represents the transition rate~\cite{Sakuraibook,newtonbook}.

The Born approximation is effected by replacing $\hat{T}$ by $\hat{V}$ and these regular matrix elements are given by $\tilde T_{\psi' \psi_{\bk}}\approx \tilde V_{\psi' \psi_{\bk}}= V_{\psi' \psi_{\bk}}/L^3$, with 
\begin{eqnarray} 
    V_{\psi' \psi_{\bk}}=\int d \br  \, e^{-i (\bk'-\bk) \cdot \br} \, \langle  \chi' \lambda' | \hat V(\br)|  \chi_{\bk \cdot \bxi} \lambda \rangle, 
\end{eqnarray}
for a local interaction $\bra{\br} \hat V \ket{\br'}=\delta(\br-\br') \hat V (\br)$. 

To compute the total probability of scattering we sum over $\lambda'$ 
and $\chi'$, and average over $\lambda$,
assuming the initial state of the target is the state $\hat\rho_{\sf t}$.
In taking the $L\rightarrow \infty$ limit,  $(L/(2\pi))^{3/2} \tilde g(\bk) \rightarrow  g (\bk)$ and $L^{-3}  \sum_\bk    \rightarrow (2 \pi)^{-3} \int d \bk$, 
and normalizing by the time-integrated average flux $I= \lim_{t \to \infty} \int_{-\infty}^t dt'  \bar{\jmath}_z(t')$, we obtain the DCS
\begin{multline}          
    \frac{d^2 \sigma}{d \Omega \, d E_{\bk'}}
    =   \frac{m^2 k'   }{16 \hbar^4 \pi^4 I} \sum_{\lambda, \lambda', \chi'} p_\lambda
    \int d k_1 \, k_1^3 d \Lambda^*_{\bk_1} d \Lambda_{\bk_2}   
    \\
    \times V^*_{\psi' \psi_{\bk_1}} V_{\psi' \psi_{\bk_2}}  
     \delta(\hbar\omega+ E_\lambda-E_{\lambda'}),
     \label{eq:main result}
\end{multline}
where the energy transfer from the constituent, incoming plane wave component to the target is $\hbar\omega \equiv E_{\bk_1}-E_{\bk'}$ and the integration measure is $d \Lambda_{\bk_i}= g(\bk_i) d\Omega_{\bk_i}$. Note that the constraint $\delta\big(\omega(k_1)-\omega(k_2)\big)$ in  Eq.~\eqref{eq:transitionprobability} enforces $k_1=k_2$. 
The probe's entanglement is encoded in  the matrix elements 
$V_{\psi' \psi_{\bk}}$, which is enhanced in magnitude whenever the relevant length scales of entanglement in the target match $\xi$. As we will see in the application (Section~\ref{sec:dimer}), there are subtle interference effects hidden in those matrix elements which are linked to entanglement. 

After scattering, the probe state becomes entangled with the target state, and the Lippmann-Schwinger equation
describes the resulting outgoing state \cite{Sakuraibook,newtonbook}. The outgoing scattered probe state in a given direction is given by 
\ignore{
\begin{eqnarray}
\hat\rho_{\sf probe}^{\sf sc}= {\sf Tr}_\lambda \left [ G_0 \hat V \hat \rho_{\sf in} \hat V^\dagger G_0^\dagger \right ],
\end{eqnarray}
where $G_0=(E_{\sf p}+E_{\lambda}-\hat H_0 +i \epsilon)^{-1}$
}
\begin{eqnarray}
\hat\rho_{\sf probe}^{\sf sc} \propto {\sf Tr}_\lambda \left [ \hat T \hat \rho_{\sf in} \hat T^\dagger \right ],\no
\end{eqnarray}
where ${\sf Tr}_\lambda$ is the partial trace over the target state space. 

A remark on the main result, Eq.~\eqref{eq:main result}, in the standard limit of plane wave scattering: when $\xi=0$ and the incident state is a plane wave normalized in a box, the flux is uniform $\bar{\jmath}_z(t) = \hbar k_0 / mL^3$. Examining the limit $L \to \infty$, we impose the concurrent restriction that $\Delta \sim L$. This ensures that the (now un-entangled) wave packet asymptotes to a plane wave while, for finite $L$, remaining square-integrable. Performing this calculation, the known form of the standard plane wave ({\sf pw}) cross section as reported e.g. in \cite{loveseybook} is recovered 
\begin{eqnarray}
    \left( \! \frac{d^2 \sigma}{d \Omega \, d E_{\bk'}} \! \right)_{\sf pw} = \mathcal{C} \frac{k'}{k_0} \sum_{\lambda, \lambda', \chi'} \! p_\lambda \left| V_{\psi' \psi_{\bk_0}} \right|^2 \delta(\hbar \omega + E_\lambda - E_{\lambda'} ) \no
\end{eqnarray}
with $\mathcal{C}= m^2 / 4 \pi^2 \hbar^4$.  

As said at the beginning of this section, we focused on the single-particle, mode-entangled probe. 
For a comparison of equivalent results when the probe is multiparticle-entangled see Appendix \ref{appendix:multiparticle}. 

\section{Magnetic Scattering}\label{sec:magnetic scattering}
%
We next extend van Hove's theory~\cite{vanhove1954,loveseybook} to the case of an entangled probe.
Consider the particular case of a neutron probe, with spin $\frac{1}{2}\hat \bsigma$ 
and mass $m=m_n$, interacting magnetically with electrons (of mass $m_e$) of the target. 
The interaction potential is given by
\begin{eqnarray}        \no
    \hat V=  \gamma \mu_N \mu_B  \sum_\j  
    \left[2\hat \bsigma \cdot \nabla \times  \frac{\hat \bs_\j \times \bR_\j}{|\bR_\j|^3}- 
    \left\{ \hat \bp_\j , \frac{\hat \bsigma \times \bR_\j}{\hbar |\bR_\j|^3} \right\}\right],
\end{eqnarray}
where $\{{\bf A} , {\bf B}\}={\bf A} \cdot {\bf B}+{\bf B} \cdot {\bf A}$, 
$\gamma=-1.913$ is the neutron's gyromagnetic ratio, $\mu_B=\frac{e \hbar}{2 m_e c}$ and 
$\mu_N=m_e\mu_B/m_n$ are the Bohr and nuclear magnetons, respectively. The 
spin and momentum of the $\j$-th electron are $\hat \bs_\j$ (eigenvalues $\pm\frac{1}{2}$) and $\hat{\bp}_{\sf j}$, and the position vectors $\bR_{\sf j}$ are directed from the $\j$-th electron
to the neutron probe. Using the identity $\bR=- \frac{R^3}{2 \pi^2}\nabla \int d \bq \frac{1}{q^2} e^{i \bq \cdot \bR}$ and completeness $\sum_{\lambda'} \ket{\lambda'}\bra{\lambda'}=\mathds 1$, evaluation of the matrix elements in Eq.~\eqref{eq:main result} yields the form
\begin{eqnarray}    
    \frac{d^2 \sigma}{d \Omega \, d E_{\bk'}}
    =   \frac{ k' r_0^2 }{4 \pi^2 I} 
    \int d k_1 \, k_1^3 d \Lambda^*_{\bk_1} d \Lambda_{\bk_2}  S(\bkappa_1, \bkappa_2, \omega) 
    \no 
\end{eqnarray}
with scattering vector $\bkappa_{1,2}=\bk_{1,2}-\bk'$, and $r_0=\frac{\gamma e^2}{m_e c^2}$. Although the two momenta, 
$\bk_{1}, \bk_{2}$, involved in this integration have the same magnitude, the variation in their direction generates a dependence of the response function $S(\bkappa_1, \bkappa_2, \omega)$ on two momentum transfers, as opposed to the single transfer 
present in a plane wave treatment. The response function is given by
\begin{eqnarray}  \label{eq:magnetic response function}    
    S(\bkappa_1, \bkappa_2, \omega)&=&\sum_\lambda \frac{p_\lambda }{2 \pi \hbar} \int_{-\infty}^\infty \! dt e^{ -i  \omega t }
       \\
    && \hspace*{-0.5cm} \times
    {\sf Tr} \big[ \hat\rho^\alpha_{\bk_1,\bk_2} \bra{\lambda}  \hat \bsigma \cdot \bQp^\dagger(\bkappa_1) \hat \bsigma \cdot \bQp(\bkappa_2, t)\ket{\lambda}  \big ] ,
    \no 
\end{eqnarray}
with magnetic interaction operator 
\begin{eqnarray}
    &&\bQp(\bkappa, t) = e^\frac{i\hat H_0t}{\hbar} \bQp(\bkappa) e^\frac{-i\hat H_0 t}{\hbar},  \,\bQp(\bkappa) =\sum_\j  \bQp^\j(\bkappa) \no \\
    &&\bQp^\j(\bkappa) =  e^{i \bkappa \cdot \mathbf{r}_\j} \left(\tilde \bkappa \times (\hat{\mathbf{s}}_\j  \times \tilde \bkappa) - \frac{i}{\hbar} \frac{\tilde \bkappa}{\kappa} \times \hat{\bp}_\j \right), \no 
\end{eqnarray}
and $\tilde \bkappa=\bkappa/\kappa$. 
The matrix $\hat{\rho}^\alpha_{\bk_1, \bk_2}=\ket{\chi_{\bk_2 \cdot \bxi}} \bra{\chi_{\bk_1 \cdot \bxi}}$ 
encodes the spin states of the entangled-probe wave packets along with the $\xi$-dependent phase-shift. Written in terms of the individual wave packets basis,
\begin{eqnarray}
    \hat \rho^\alpha_{\bk_1,\bk_2} = \frac{1}{2}\sum_{\nu,\nu'} e^{i \frac{(-1)^\nu \Theta_{\bk_1}-(-1)^{\nu'} \Theta_{\bk_2} }{2} } \ket{\chi^\alpha_{\nu'}} \bra{ \chi^\alpha_{\nu}},
\end{eqnarray}
whose trace is given by 
\begin{eqnarray}
    {\sf Tr} [\hat \rho^\alpha_{\bk_1,\bk_2}]=\cos \left(\frac{\bk_1-\bk_2}{2} \right)\! \cdot\! \bxi ,
\end{eqnarray}
and it is a pure-state density matrix 
when $\bk_1=\bk_2$: $\hat \rho^\alpha_{\bk_1}\equiv\hat \rho^\alpha_{\bk_1, \bk_1}=(\mathds{1}+\hat \bsigma \cdot \hat \chi_\alpha)/2$. 

Using this formalism, the trace in Eq.~\eqref{eq:magnetic response function} gives a decomposition of $S(\bkappa_1, \bkappa_2, \omega)$ into the four combinations of spin components involved, 
\begin{eqnarray} \nonumber
    S(\bkappa_1, \bkappa_2, \omega)=\frac{1}{2}\sum_{\nu,\nu'} e^{i\frac{(-1)^\nu \Theta_{\bk_1}-(-1)^{\nu'} \Theta_{\bk_2} }{2} } S_{\nu\nu'}(\bkappa_1, \bkappa_2, \omega) ,
\end{eqnarray}
so that spin-diagonal ($\nu = \nu'$) entries describe scattering due to each individual wave packet of the probe while the off-diagonal ($\nu \neq \nu'$) entries describe interference between them.

Similarly one can derive the polarization vector $\mathbf{P}'$~\cite{loveseybook}  of the scattered neutron, defined as 
\begin{eqnarray}
	\mathbf{P}' &=& \frac{ {\sf Tr} [\hat{\rho}^{\sf sc}_{\sf probe} \hat{\bsigma}]}{ {\sf Tr} [\hat{\rho}^{\sf sc}_{\sf probe}] }, \no
\end{eqnarray}

to be given by 
\begin{eqnarray}    \label{Eq:Polarization}
     &&  \left( \frac{d^2 \sigma}{d \Omega \, d E_{\bk'}} \right) {\mathbf P'} 
    =   \frac{ k' r_0^2 }{8 \pi^3 \hbar I} 
    \int d k_1 \, k_1^3 d \Lambda^*_{\bk_1} d \Lambda_{\bk_2} 
     \sum_\lambda p_\lambda  \\
    && \int_{-\infty}^\infty \! dt e^{ -i  \omega t }
    {\sf Tr} \big[ \hat\rho^\alpha_{\bk_1,\bk_2} \bra{\lambda}  \hat \bsigma \!\cdot \! \bQp^\dagger(\bkappa_1) \, \hat \bsigma \ \hat \bsigma \ \!\cdot  \! \bQp(\bkappa_2, t)\ket{\lambda}  \big ] .
    \no 
\end{eqnarray}

\begin{figure}[thb]
    \centering
    \includegraphics[width=\columnwidth]{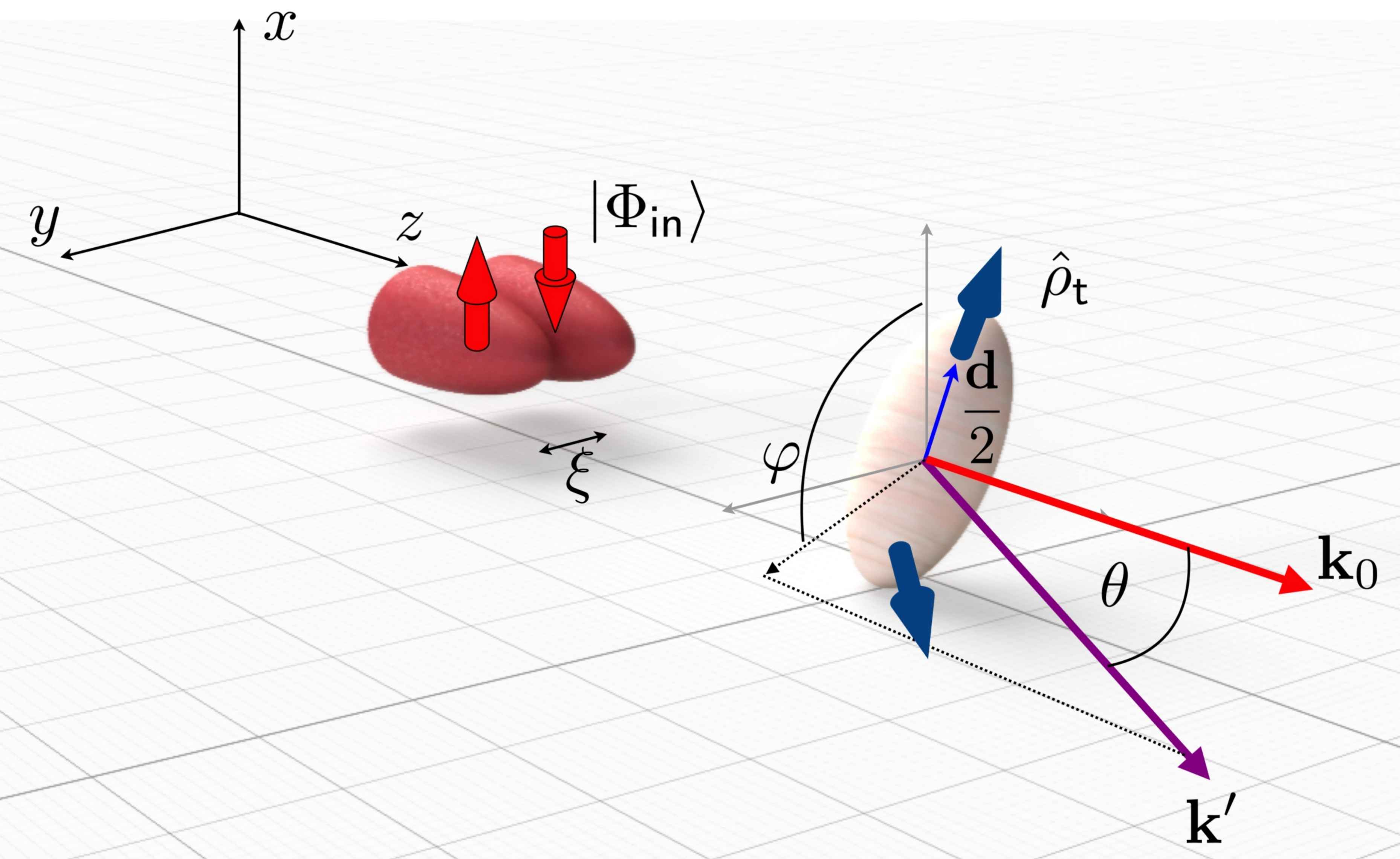}
    \caption{Magnetic scattering of an entangled-probe with entanglement length $\xi$, by a dimer of size $|\bd|$, into scattering angles $(\theta,\varphi)$ for a general orientation of the dimer. The spin-quantization axis for the probe is ${\alpha=x}$, i.e., ${\ket{\!\!\uparrow}_x}$,$\ket{\!\!\downarrow}_x$. The initial state of the probe-target system is $\hat \rho_{\sf in}=\ket{\Phi_{\sf in}}\bra{\Phi_{\sf in}}\otimes \hat \rho_{\sf t}$.}
    \label{fig:dimer orientation general}
\end{figure}

\section{Uncovering Entanglement from Entanglement}\label{sec:dimer}

\begin{table*}[t]
\begin{centering}
\renewcommand{\arraystretch}{1.3}
\begin{tabular}{|c|c||c|c||c|c|}
    \hline
    & $\tau \rightarrow \tau'$ & $A_{\tau \tau'} $ & ${\bf B}_{\tau \tau'}$ & $\tilde A_{\tau \tau'} $ & $\tilde {\bf B}_{\tau \tau'}$\\
    \hline
    \hline
{Any $\sf T$}    & $s \rightarrow t$ & $1+(\tilde{\bkappa}_1 \cdot \tilde{\bkappa}_2)^2$ & $(\tilde{\bkappa}_1 \cdot \tilde{\bkappa}_2)\tilde{\bkappa}_1 \times \tilde{\bkappa}_2$ & 2 & 0\\
    \hline 
    \multirow{2}[2]*{${\sf T}=0$} 
    & $t\rightarrow s$ & $\begin{aligned} 1+&(\bc^* \cdot \tilde{\bkappa}_1)(\bc\cdot \tilde{\bkappa}_2 )( \tilde{\bkappa}_1 \cdot \tilde{\bkappa}_2) \\ -& (\bc \cdot \tilde{\bkappa}_1) (\bc^* \cdot \tilde{\bkappa}_1 ) - (\bc \cdot \tilde{\bkappa}_2 )( \bc^* \cdot \tilde{\bkappa}_2) \end{aligned}$ & $ \begin{aligned} \bc^* &\times \bc + (\bc^* \cdot \tilde{\bkappa}_1) (\bc \cdot \tilde{\bkappa}_2 )( \tilde{\bkappa}_1 \times \tilde{\bkappa}_2) \\ &+ (\bc^* \cdot \tilde{\bkappa}_1)(\bc \times \tilde{\bkappa}_1) - (\bc \cdot \tilde{\bkappa}_2)(\bc^* \times \tilde{\bkappa}_2) \end{aligned} $ & $1-|\tbkappa \cdot \bc |^2$ & $\tbkappa [ \tbkappa \cdot (\bc^* \times \bc)]$\\
    \cline{2-6}
    & $t\rightarrow t$ & $A_{st} - A_{ts}^*$ & $\bB_{st} - \bB_{ts}^*$ & $1+|\tbkappa \cdot \bc |^2$ & $\tbkappa [ \tbkappa \cdot (\bc^* \times \bc)]$\\
    \hline 
    \multirow{2}[0]*{${\sf T}>0$} 
    & $t \rightarrow s$ & $A_{st}$ & $ \bB_{st} $ & $2$ & $0$\\
    \cline{2-6}
    & $t \rightarrow t$ & $2 A_{st} $ & $2 \bB_{st} $ & $4$ & $0$\\
    \hline 
\end{tabular}
\end{centering}
    \caption{Variables entering Eqs.~\eqref{eq:htautau} and \eqref{eq: Response function_PW} for zero and finite temperature, $\sf T$, which plays the role of adjusting the singlet/triplet admixture of the thermal state as described in the text.  The normalized scattering vector is $\tilde{\bkappa}= \bkappa/\kappa$, $s$ and $t$ stand for {\it singlet} and {\it triplet}, respectively, and $\tilde A_{\tau \tau'}$ and $\tilde {\bf B}_{\tau \tau'}$ represent the plane-wave limit of $A_{\tau \tau'}$ and ${\bf B}_{\tau \tau'}$. }
    \label{table:1}
\end{table*}

To highlight the information entangled-probe scattering can provide, we apply our framework to the 
case of a target with two motionless interacting electrons, i.e., a dimer. Electrons are positioned at $\br_\j=(-1)^\j\bd/2$, $\j=0,1$ (Fig.~\ref{fig:dimer orientation general}), and their interaction is governed by the Heisenberg Hamiltonian $\hat{H}_{\sf t}=-4 J ~ \hat \bs_0 \cdot \hat \bs_1$ with exchange coupling $J$. 
Its Hilbert space is the direct sum of singlet and triplet subspaces: $\cH_{\sf target}=\cH_s\oplus \cH_t$, where 
\begin{eqnarray}
    \cH_s &=& {\sf Span}\Big\{\ket{\lambda_s}=\frac{1}{\sqrt 2}(\ket{\up \down}_z-\ket{\down \up}_z)\Big\}, \no \\
    \cH_t &=& {\sf Span}\Big\{\ket{\lambda_x}, \ket{\lambda_y}, \ket{\lambda_z}\Big\}  \quad , \mbox{with} \, \ket{\lambda_\a}= (-1)^\j 2 \hat s_{\j}^\a \ket{\lambda_s} ,\no 
\end{eqnarray}
where $\ket{\!\! \uparrow}_\alpha$, $\ket{\!\!\downarrow}_\alpha$ are spin eigenstates defined along the $\alpha$ spin-quantization axis. More specifically, 
\begin{gather}		\no
		\ket{\lambda_x}=-\frac{\ket{\up\up}_z-\ket{\down \down}_z }{\sqrt{2}}, \quad 
    \ket{\lambda_y}=i \frac{ \ket{\up\up}_z+\ket{\down \down}_z }{\sqrt{2}}, 
    \\ 
    \ket{\lambda_z}=\frac{\ket{\up\down}_z+\ket{\down \up}_z }{\sqrt{2}} .
    \no
\end{gather}
These are energy eigenstates, i.e., $\hat{H}_{\sf t}\ket{\lambda_s}=E_{\lambda_s}\ket{\lambda_s}$ and 
$\hat{H}_{\sf t}\ket{\lambda_\alpha}=E_{\lambda_t}\ket{\lambda_\alpha}$ with 
$E_{\lambda_s} = 3J$ and $E_{\lambda_t} = -J$. 

A standard physical measure of multipartite entanglement is the purity \cite{barnum-2004, barnum-2003, somma-2004}, a pedagogical explanation of which is found in~\cite{OrtizChapter}. Given a normalized state $\ket{\lambda_t}=\sum_\a c_\a \ket{\lambda_\a} \in \cH_t$, its purity is given by
\begin{eqnarray}
    P_{su(2)\oplus su(2)}(\ket{\lambda_t})=2\sum_{\alpha,\j} \bra{\lambda_t}\hat s_{\j}^\a \ket{\lambda_t}^2=|{\bf c}^* \times {\bf c}|^2 \no 
\end{eqnarray}
with $\bc=(c_x, c_y, c_z)$ encoding the coefficients of the linear combination. A pure triplet state is maximally entangled (un-entangled) if and only if $\bc^* \times \bc={\bf 0}$, i.e. $\bc $ is a real-valued vector ($|\bc^* \times \bc|=1$). An example of a state exhibiting partial entanglement is
\begin{eqnarray}
	\ket{\lambda_t} &=& \frac{ \ket{\!\! \uparrow \uparrow}_z + \sqrt{3} \ket{\!\! \downarrow \downarrow}_z }{2}, \no
\end{eqnarray}
which has coefficient vector ${\mathbf{c}= \left( \frac{\sqrt{3}-1}{2 \sqrt{2}} , -i \frac{\sqrt{3} + 1}{2 \sqrt{2}} , 0 \right) }$ and purity of $1/4$.
This quantification allows for a simple identification of entanglement-induced features of the calculated cross section, as there will be terms proportional to the purity which then vary depending on the degree of entanglement. 

\begin{figure}[t]
    \centering
    \includegraphics[height=0.6\columnwidth]{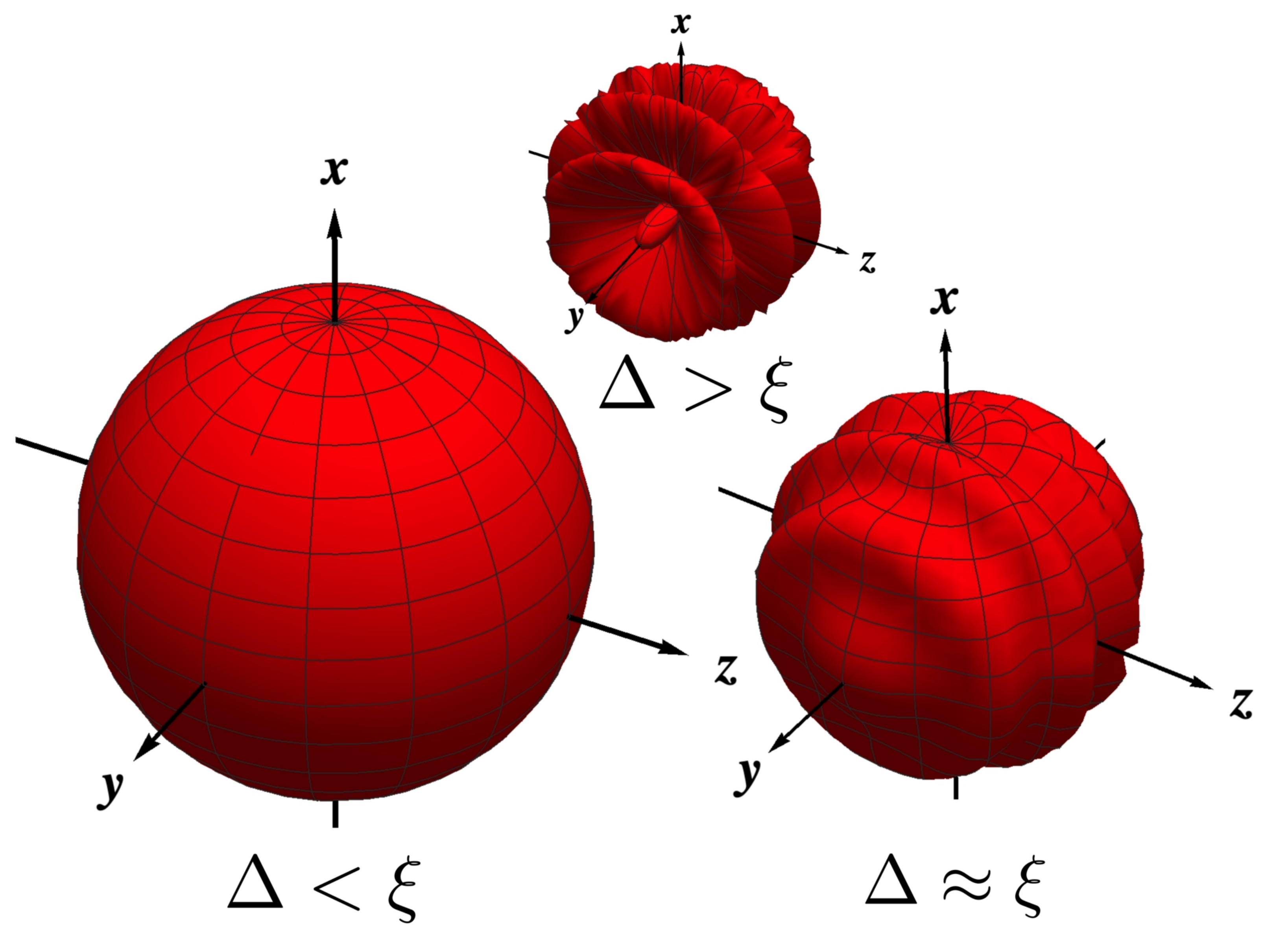}
    \caption{Thermal state $\hat \rho_{\sf t}$ triplet-to-singlet DCSs $d^2 \sigma/d \Omega \, d E_{\bk'}\big|_{t\to s}$ (spherical plots $(\theta,\varphi)$) 
    for the case $|\bd_\perp|\approx \xi$, $\phi=0$, $k_0\approx 1.5\times 10^4 \mu{\rm m}^{-1}$, and the dimer aligned along the $y$ direction. 
    The DCS for $\Delta < \xi$ is proportional to $\sum_\alpha (1-|\tilde \bkappa_0 \dot \bc|^2)=2$, with $\bc$ corresponding to $\ket{\lambda_\alpha}$.
    While the ordering of 
    sizes is faithful, plots do not have the same scale for visualization purposes, as increases in $\Delta$ achieve sharp decreases in the DCS magnitudes when 
    $|\bd_\perp|\approx \xi$.}
    \label{fig:crosssection}
\end{figure}

\begin{figure}[htb]
    \centering
    \includegraphics[height=0.8\columnwidth]{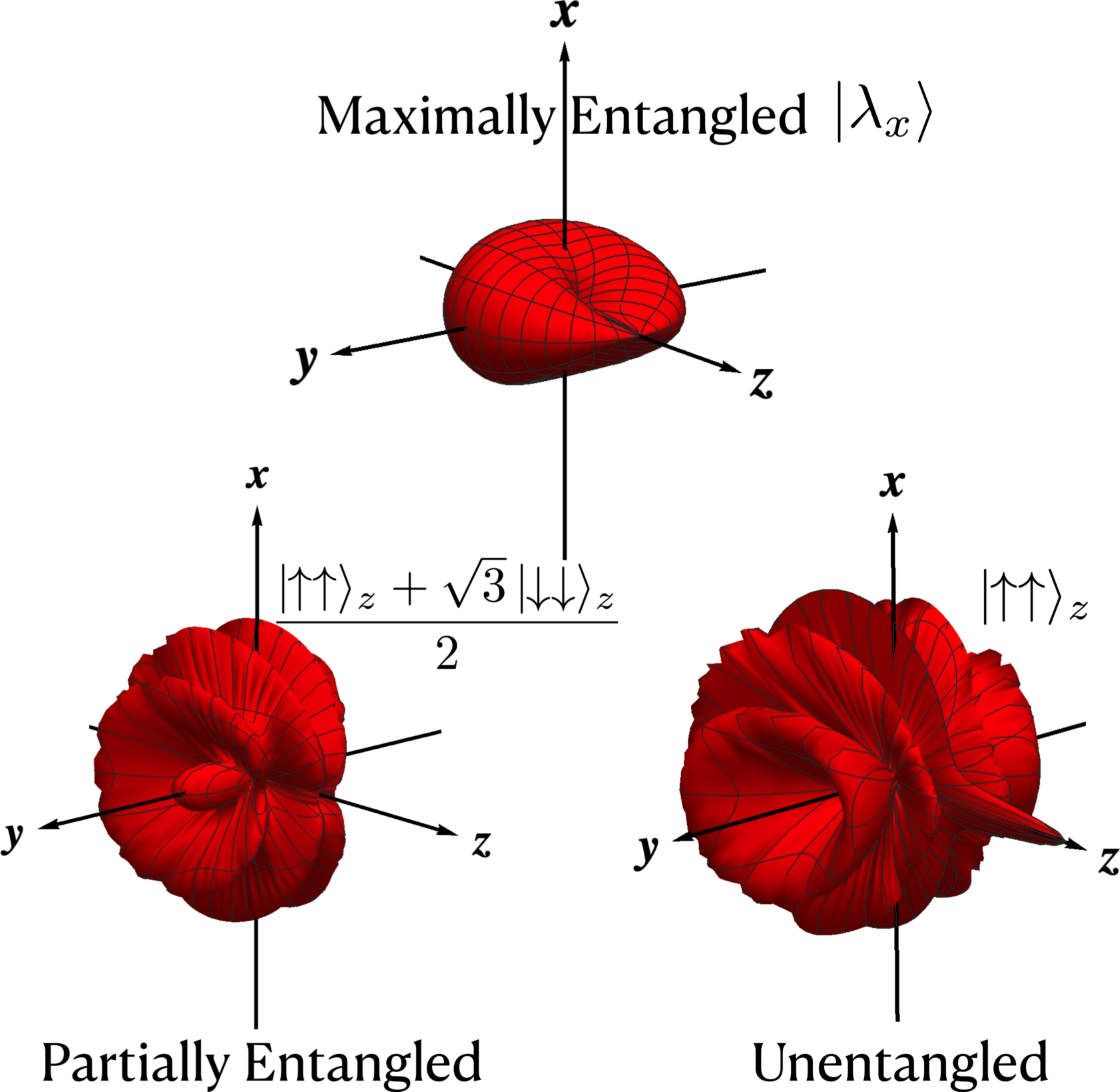}
    \caption{Pure state triplet-to-singlet DCSs  $d^2 \sigma/d \Omega \, d E_{\bk'}\big|_{t\to s}$ (spherical plots $(\theta,\varphi)$) for the case 
    $|\bd_\perp|\approx \xi$, $\Delta < \xi$, $\phi=0$, $k_0\approx 1.5\times 10^4 \mu{\rm m}^{-1}$, and the dimer aligned along the $y$ direction. 
    DCSs are to scale with each other, illustrating the effect target-state entanglement has on the DCS interference pattern. The DCS for $\ket{\lambda_\alpha}$ 
    is proportional to $1-|\tilde \bkappa_0 \dot \bc |^2$. Purity values for the displayed maximally-, partially- and un-entangled states are 0, 1/4, and 1, respectively.}
    \label{fig:triplet to singlet}
\end{figure}

The initial state of the entangled probe is defined by 
a Gaussian distribution 
\begin{eqnarray}
    g(\bk)=\left(\frac{\D}{ \sqrt{2 \pi}}\right)^{\frac{3}{2}} e^{-\frac{\D^2}{4}|\bk-\bk_0|^2}, \no
\end{eqnarray} 
%
with average momentum $\bk_0=k_0 \hat z$ ($k_0\approx 1.5\times 10^4 \mu{\rm m}^{-1}$ in \cite{shen2019unveiling,kuhn2020unveiling}), 
spatial width $\D$, spin-quantization axis ${\alpha=x}$, and tunable entanglement vector $\bxi$ in the $y$-direction ($25 nm <\xi< 25 \mu{\rm m}$ in 
\cite{shen2019unveiling,kuhn2020unveiling}, see Fig.~\ref{fig:dimer orientation general} for the general setup and Fig.~\ref{fig:dimer orientation specific} for the specific setup used in the the plots of Figs.~\ref{fig:crosssection} and \ref{fig:triplet to singlet}). 
We consider two types of initial target states $\hat \rho_{\sf t}$: 
\begin{eqnarray}
  \hat \rho_{\sf t} &=& p_s \ket{\lambda_s}\! \bra{\lambda_s} + p_t \, \sum_\alpha \ket{\lambda_\alpha} \bra{ \lambda_\alpha } \quad \mbox{(Thermal)}, \no \\
  \hat \rho_{\sf t}&=& \ket{\lambda_s}\! \bra{\lambda_s} \ \  \mbox{or}  \ \ \ket{\lambda_t} \bra{\lambda_t} \quad \quad \quad \mbox{(Pure State)} \no
\end{eqnarray}
($p_{s,t}\ge 0$, ${\sf Tr}_{\lambda}[ \hat \rho_{\sf t} ]=1$).  In the thermal state, $p_s=e^{-\frac{3J}{k_B {\sf T}}}/{\cal Z}$ and 
$p_t=e^{\frac{J}{k_B {\sf T}}}/{\cal Z}$, with ${\cal Z}=e^{-\frac{3J}{k_B {\sf T}}}+3 e^{\frac{J}{k_B {\sf T}}}$, are Boltzmann factors which incorporate 
the effect of temperature ${\sf T}$ in the scattering cross section. Investigation of its pure state components will permit analysis of the effect of the target's entanglement on the DCS. 

The computed total response function is a sum of three components based on the type of transition that occurs in the target,
\begin{eqnarray} 
    S(\bkappa_1, \bkappa_2, \omega) &=& p_s S_{s\rightarrow t}(\bkappa_1, \bkappa_2, \omega) \no \\
    && \hspace{-0.7cm} + p_t \Big(S_{t\rightarrow s}(\bkappa_1, \bkappa_2, \omega) + S_{t\rightarrow t}(\bkappa_1, \bkappa_2, \omega)\Big). \quad \no
\end{eqnarray}
These terms can be factored in such a way as to isolate the information pertaining to the dimer from that of the entanglement of the probe ($k_1=k_2$): 
\begin{eqnarray}\hspace*{-0.4cm}
    S_{\tau \to \tau'} (\bkappa_1, \bkappa_2, \omega) =  \delta(\hbar\omega +4 J \zeta_{\tau  \tau'} ) F_{\tau \tau'} (\bd) h_{\tau \tau'}(\bxi),
\end{eqnarray}
with  $\zeta_{st} = +1$, $\zeta_{ts} = -1$, and $\zeta_{tt} = 0$. Conservation of energy implies 
${k'}^2=k_1^2-\frac{2 m \omega}{\hbar}=k_1^2+\frac{8 m J \zeta_{\tau  \tau'}}{\hbar^2}$.
In the above decomposition we have introduced real functions describing the dimer structure, 
\begin{eqnarray}
     F_{\tau \tau'}(\bd) &=& 2 \cos \left( \frac{\bkappa_1 - \bkappa_2}{2} \right) \cdot \bd \no \\
    && \hspace{0.5cm} - (-1)^{\delta_{\tau \tau'}} 2 \cos \left( \frac{\bkappa_1 + \bkappa_2 }{2} \right) \cdot \bd , \no
\end{eqnarray}
and functions encoding the entanglement length of the probe, 
\begin{eqnarray}
   \hspace*{-0.6cm} h_{\tau \tau'}(\bxi) \!&=&\! A_{\tau \tau'} \cos \frac{\Theta_{\bk_1}\!\! -\! \Theta_{\bk_2}}{2} \!+\! i {\bf B}_{\tau \tau'}\! \cdot\! \bra{\chi_{\bk_1\cdot\bxi}} \hat \bsigma \ket {\chi_{\bk_2\cdot\bxi}}   . \label{eq:htautau}
\end{eqnarray}
The expressions for $A_{\tau \tau'}$ and ${\bf B}_{\tau \tau'}$ are summarized in Table~\ref{table:1}.
The entanglement length enters into $h_{\tau \tau'}$ via the first term as well as the $\bxi$-dependence of the matrix element
\begin{eqnarray}
    \bra{\chi_{\bk_1\cdot\bxi}} \hat \bsigma \ket {\chi_{\bk_2\cdot\bxi}} && \no \\
      && \hspace*{-2cm}= \! \bigg( \! i \sin  \frac{\Theta_{\bk_1}\! -\! \Theta_{\bk_2}}{2}  ,\! -\! \sin   \frac{\Theta_{\bk_1}\!+\!\Theta_{\bk_2}}{2}    ,  \cos   \frac{\Theta_{\bk_1}\!+\!\Theta_{\bk_2} }{2}     \bigg) . \no
\end{eqnarray}
\begin{figure}[htb]
    \centering \hspace*{-0.3cm}
    \includegraphics[height=0.77 \columnwidth]{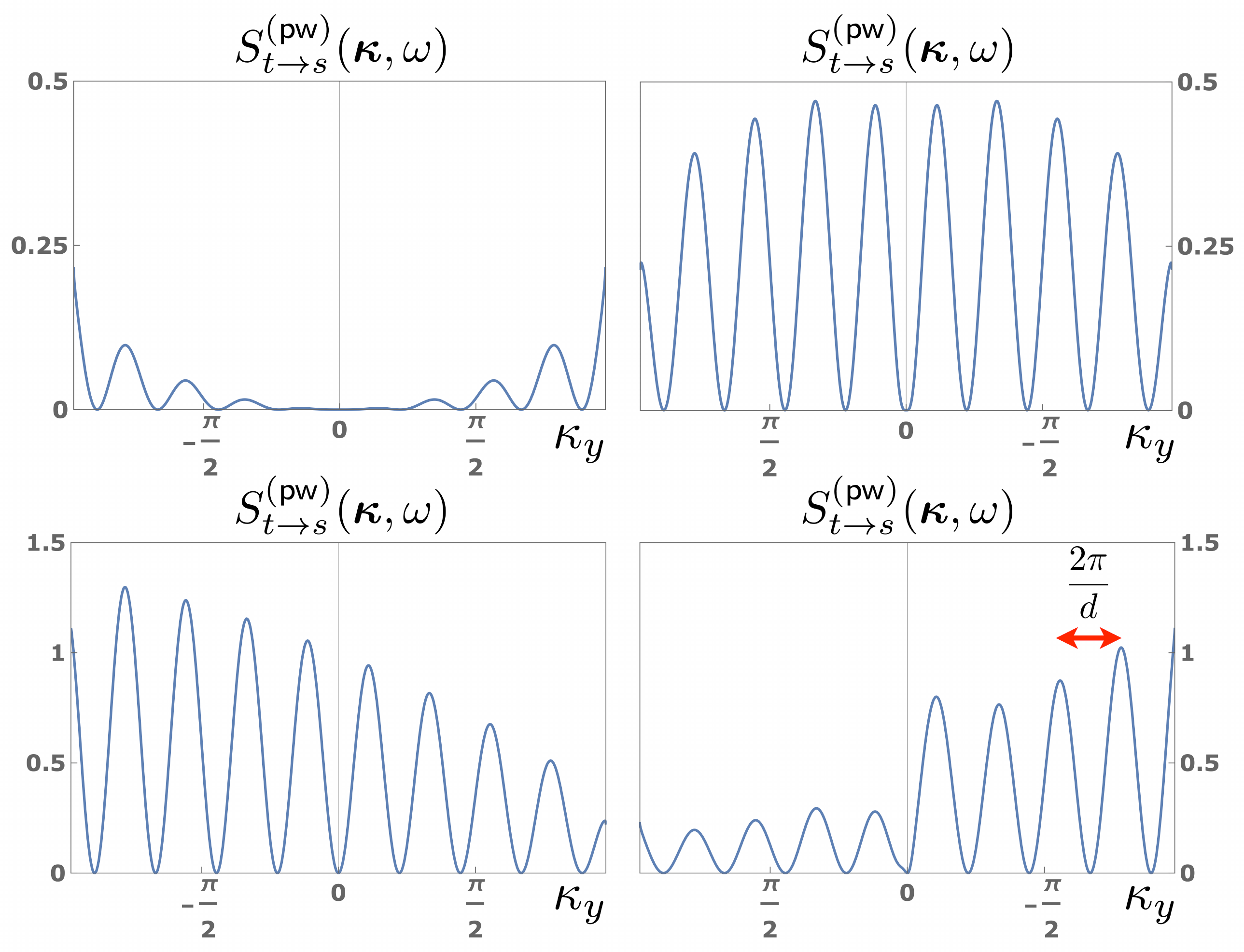}
    \caption{Magnetic response function \eqref{eq: Response function_PW}, i.e., plane-wave limit, for the triplet $\ket{\lambda_t}=\ket{\up \up}_z$ to singlet $\ket{\lambda_s}$ transition,  $S^{({\sf pw})}_{t \to s}(\bkappa, \omega)$, when $\kappa_x=0$ assuming 
    $k=\pi${\AA}$^{-1}$, $d=9${\AA},  $J=1/4$ meV, and ${k'}^2=k^2-\frac{8m J}{\hbar^2}$. 
    Top panels depict the case $\xi=0$, corresponding to an incoming neutron of spin polarization in the 
    $+z$ direction, with forward (backward) scattering $0\le \theta < \pi/2$ ($\pi/2\le \theta \le \pi$) on the right (left). Bottom panels display the same information when, by tuning $\bxi$ ($\bk \! \cdot \! \bxi = 3 \pi/2$, $\phi=0$), the effective polarization is set along the $+y$ direction. }
    \label{fig:Plane-wave-limit}
\end{figure}

There are three competing length scales in the problem: $|\bd_\perp|$, that is the projection of $\bd$ 
onto the $x$-$y$ plane, $\xi$, and $\Delta$ (axes and dimer orientation coincide with 
Fig.~\ref{fig:dimer orientation specific}). To gain intuition into the kind of additional information entangled-beam scattering provides, we start by considering 
the magnetic response function of the dimer 
in the limit $\Delta \rightarrow \infty$,  $\bk_1=\bk_2=\bk=k \hat z$, and $\xi \ll \Delta$ can be different from $0$. 
As $\Delta$ eclipses $\xi$, the wave packets overlap creating a wave function which is almost indistinguishable from the 
unentangled plane wave, but which still has a technical $\bxi$ dependence in the form of a phase. For this reason, 
we refer to this limit as the plane-wave ({\sf pw}) limit.
 As for the entangled wave packet probe, the total response function can be decomposed into a sum of three components, 
\begin{eqnarray} 
    S^{(\sf pw)}(\bkappa, \omega) &=& p_s S^{(\sf pw)}_{s\rightarrow t}(\bkappa, \omega) \no \\
  &&+ \, p_t \Big(S^{(\sf pw)}_{t\rightarrow s}(\bkappa, \omega) + S^{(\sf pw)}_{t\rightarrow t}(\bkappa, \omega)\Big),   \no
\end{eqnarray}
where $\bkappa=\bk-\bk'$ is the momentum transfer, and
\begin{eqnarray}
      S^{(\sf pw)}_{\tau \to \tau'} (\bkappa, \omega)
    &=&\delta(\hbar \omega +4 J \zeta_{\tau  \tau'} )
     \sin^2 \left( \frac{\bkappa \!\cdot\! \bd+\pi \delta_{\tau \tau'}}{2}\right)  
    \no     \\
    && \hspace*{0.5 cm} \times  \big[ 
    \tilde A_{\tau \tau'} + i \tilde \bB_{\tau \tau'} \cdot  \hat{\chi}_x
     \big],
    \label{eq: Response function_PW}
\end{eqnarray}
where expressions for $\tilde A_{\tau \tau'}$ and $\tilde \bB_{\tau \tau'}$ take the simpler form shown in
Table~\ref{table:1} ($\tilde \bB_{\tau \tau'}$ is purely imaginary).  The term $\sin^2 ( \frac{\bkappa \cdot \bd+\pi \delta_{\tau \tau'}}{2})$ is typical 
of a two slit-type interference pattern, with the dimer playing the role of the slits. 

Although the initial probe is in the plane-wave limit, there is still in principle a 
dependence of the scattered state on the path entanglement vector $\bxi$. Notice, however, that 
when the target state is maximally entangled its purity $|\bc^* \times \bc|^2$ vanishes and, since 
$\tilde \bB_{\tau \tau'}$ is directly proportional to the purity, the magnetic 
response $S^{(\sf pw)}_{\tau \to \tau'}$ is insensitive to $\xi$. Thus, maximal entanglement of the target in the plane wave case precludes $\bxi$-dependence of the DCS -- a dependence which {\it is} present for nonzero purity (see Fig. \ref{fig:Plane-wave-limit}). 

Interestingly, the role $\Theta_\bk$ plays in the response is 
effectively equivalent to a rotation of the spin polarization of the neutron. In other words, in this 
plane-wave limit and from the standpoint of the DCS, tuning the properties of the beam 
by manipulation of the entangler is similar to changing the polarization of the incident neutrons. 
We next analyze the polarization of the scattered neutron in the plane-wave limit. From Eq. \eqref{Eq:Polarization},
\begin{eqnarray} 
	 \bP'_{\tau \to \tau'}	
    =\frac{1}{ \tilde A_{\tau \tau'}  + i \tilde {\bf B}_{\tau \tau'} \dot \hat{\chi}_x}  \tilde \bh_{\tau \tau'}(\bxi),
\end{eqnarray}
with
\begin{eqnarray}
	\tilde \bh_{st}(\bxi) &=&   -  2 \tbkappa \, \hat{\chi}_x \dot \tbkappa, 
	\hspace*{2.5 cm}\mbox{for any} \quad \!\! \! {\sf T} 
	\no 		\\
	\tilde \bh_{t \tau'}(\bxi) 
    &=&  (-1)^{\d_{t \tau'}}\Re \Big[   2   \bc_\perp \, \hat{\chi}_x \dot \bc_\perp^*  
	 -  \hat{\chi}_x  \, c_\perp^2 \Big]   - i \tilde {\sf \bB}_{t \tau'}  
	 \no		\\	  
    &&+ \d_{t \tau'} \tilde \bh_{st}(\bxi) , \hspace*{2.15 cm} \mbox{for}
    \quad {\sf T}=0 
    \no	\\
    \tilde \bh_{t\tau'}(\bxi) &=&   \d_{t \tau'} \tilde \bh_{st}(\bxi), \hspace*{2.45 cm}
     \mbox{for}       \quad {\sf T}>0,
    \no 
\end{eqnarray}
and $\bc_\perp= \bc  -\tbkappa \, ( \tbkappa \dot \bc)$.
For example, if the incident polarization is in the $y$-$z$ plane, i.e. $\hat \chi_x$, and we 
restrict $\bk'$ to the $y$-$z$ plane ($\tilde \k_x=0$), then the polarization of the scattered neutron for the triplet-to-singlet transition at ${\sf T} =0$ is
\begin{eqnarray}
	\bP'_{t\to s } (\lambda_x \to \lambda_s)&=& -  \hat \chi_x, 
	\no	\\
	\bP'_{t\to s } (\lambda_y \to \lambda_s)&=& \frac{2}{c^2_\perp} \bc_\perp \hat \chi_x \dot \bc_\perp  \! - \! \hat \chi_x=	\bP'_{t\to s } (\lambda_z \to \lambda_s),
	\no
\end{eqnarray}
with  $ \bc_\perp/ c_\perp= (0,\tilde \k_z, - \tilde \k_y)$ and  $c_\perp^2=1 -|\tilde \bkappa \dot \bc|^2$.
The situation becomes more interesting away from the plane-wave limit as discussed next. 

\begin{figure}[hbt]
    \centering
    \includegraphics[height=0.6\columnwidth]{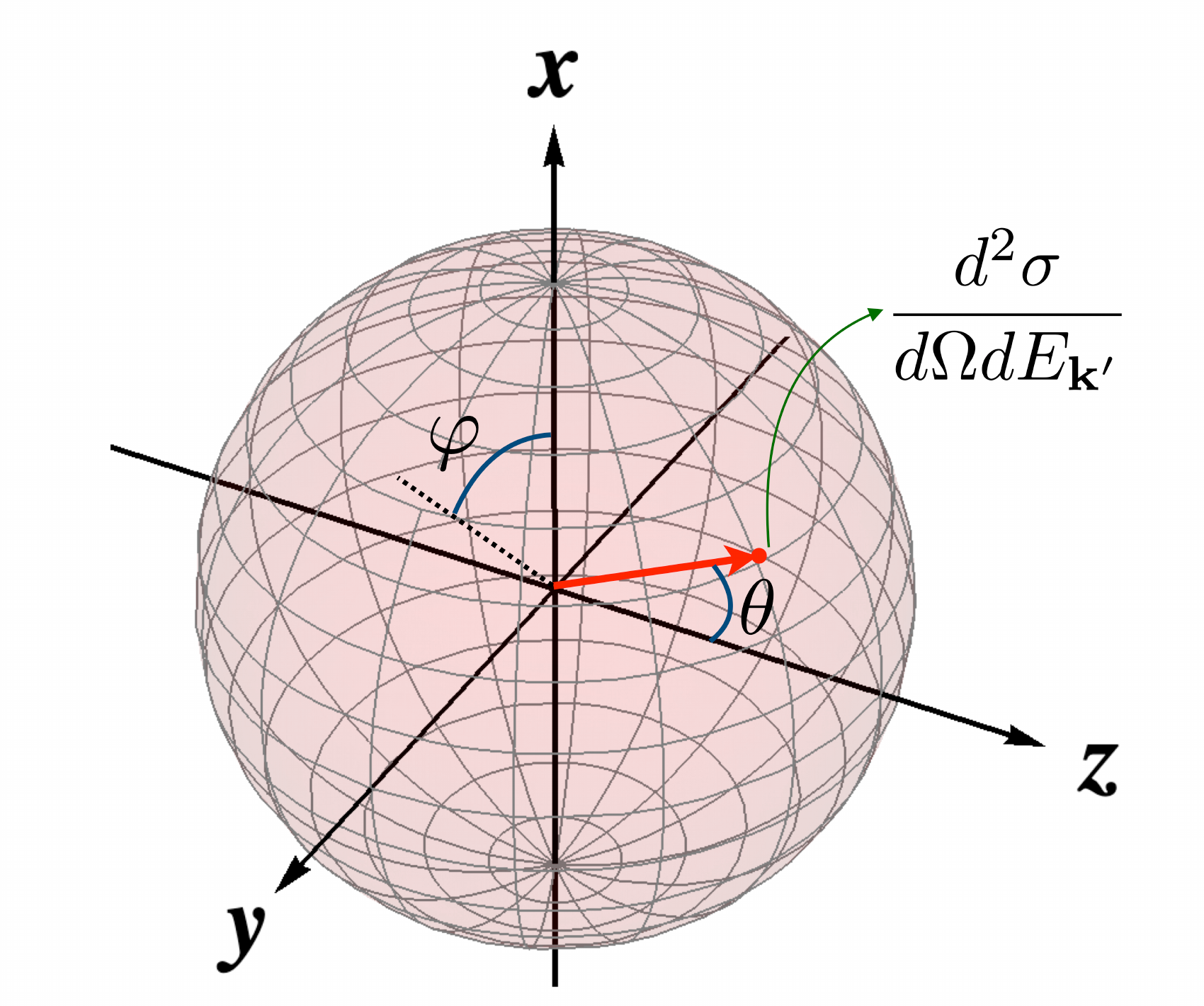}
    \caption{The distance from the origin to the surfaces in Figs. ~\ref{fig:crosssection} 
    and~\ref{fig:triplet to singlet} represent the magnitude of the DCS as a function of scattering direction, $\bk'$ (defined by the angles $\theta$ and 
    $\varphi$).}
    \label{fig:SphericalPlotSC}
\end{figure}

Returning to the main focus of the entangled probe, the DCS also contains three components: 
\begin{eqnarray}		\no
	\frac{d^2 \sigma}{d \Omega \, d E_{\bk'}}
	\!=p_s \frac{d^2 \sigma}{d \Omega \, d E_{\bk'}}\bigg|_{s\to t}
	\!\!\!\!\!\!+p_t \frac{d^2 \sigma}{d \Omega \, d E_{\bk'}}\bigg|_{t\to s}
	\!\!\!\!\!\!+p_t\frac{d^2 \sigma}{d \Omega \, d E_{\bk'}}\bigg|_{t\to t}\!\!.
\end{eqnarray}
Figures~\ref{fig:crosssection} and~\ref{fig:triplet to singlet} show the shape of the DCS for various parametric scenarios. 
The radial value of these plots in the direction $(\theta, \phi)$ gives the value of $\frac{d^2\sigma}{d\Omega dE_{\bk'}}$ in 
that direction, as illustrated in Fig.~\ref{fig:SphericalPlotSC}. The axes and dimer orientation coincide with Fig.~\ref{fig:dimer orientation specific}. 
\ignore{The largest scattering amplitudes occur when 
$|\bd_\perp|\sim \xi$ or when $\Delta$ is the 
dominant scale (i.e., $\Delta > |\bd_\perp| , \xi$ and $|\bd_\perp| \ne \xi$).} 

The largest scattering amplitudes occur when $|\bd_\perp| \sim \xi$ and $\Delta$ is small. As soon as the value $|\bd_\perp|$ departs from the 
length $\xi$, the DCS starts to attenuate exponentially, at which point a large $\Delta$ is required to counter this effect.  
Figure~\ref{fig:crosssection} compares the angular dependence of DCSs in a triplet-to-singlet transition of a thermal initial 
target state $\hat{\rho}_{\sf t}$ for different competing length scales and $\tbkappa_0=(\bk_0\!-\!\bk')/|\bk_0\!-\!\bk'|$. In each of these, 
$\xi \approx | \bd_\perp| \neq 0$ and so the beam is entangled. 
The case $\Delta > \xi$ 
does not differ qualitatively from what one would have obtained with 
an un-entangled probe ($\xi=0$). Indeed, the ``flower-shape'' DCS when $\Delta > \xi$ is reminiscent of a two-slit-type interference pattern  
and, as mentioned above, is also obtained in the case of a standard probe ($\xi=0$).
On the other hand, the behavior of a non-overlapping entangled 
wave packet ($\D < \xi \sim |\bd_\perp|\ $) is different than an un-entangled probe: the distinct two-slit interference pattern vanishes, 
leaving a DCS which is insensitive to $(\theta, \phi) \in [ 0,\pi ] \times [ 0,2\pi )$ and far stronger in magnitude. This spherical symmetry 
arises from a delicate summation of contributions from the triplet states making up $\hat{\rho}_{\sf t}$, which do not individually exhibit 
this feature (for example, see the $\ket{ \lambda_x}$ DCS of Fig.~\ref{fig:triplet to singlet}).

The situation is even more remarkable when the target state is pure. Then, some interference terms 
are proportional to the purity $P_{su(2)\oplus su(2)}(\ket{\lambda_t})$ of the target state and 
consequently the DCS can identify entanglement in the target. Figure \ref{fig:triplet to singlet} 
displays the triplet-to-singlet DCS for three particular target states using the same entangled, $\Delta < \xi$ neutron probe.
Maximally entangled Bell-type states of the target show a special shape distinct from those of un-entangled or 
partially entangled states. The latter depict two-slit-type interference patterns with proper characteristics of the particular symmetry of the 
probe-target system, while Bell-type target states seem to forbid those two-slit-type interference patterns 
 as a result of their non-local correlations, i.e., their entanglement.
  Such  ``quantum erasure" of the  interference pattern can be understood by realizing that orthogonality of the incident spin states 
  corresponding to the two paths are preserved after scattering if and only if the  target state is maximally entangled.  Let us expand:
  
In the limit  $\D\ll \xi,|\bd_\perp|$, the neutron wave packet scatters significantly only from those scattering centers that lie in its trajectory, i.e. the dimer spin 
$\hat s_\j$ interacts mainly with the neutron spin state $\ket{\chi^\a_\j}$. Consider, for instance, the scattered neutron spin-dimer 
triplet-to-singlet transition 
\begin{eqnarray}
	\ket{\chi^{\sf sc}_\j}\ket{\lambda_s}
	=\ket{\lambda_s}\!\bra{\lambda_s} \big[\bQp^\j\dot \hat \bs_\j \ket{\chi^\alpha_\j} \ket{\lambda_t}\big]
	=e^{i \bkappa_0 \dot \br_\j} \hat \bsigma \dot \bc_\perp \ket{\chi^\a_\j} \ket{\lambda_s},
	\no
\end{eqnarray}
where $\ket{\chi^{\sf sc}_\j}$ is the spin state corresponding to the $\j$-th neutron wave packet after scattering with 
final momentum $\bk'$, and $\bc_\perp= \bc  \!-\!\tbkappa_0 \, ( \tbkappa_0 \dot \bc)$. 
The path interference term in the DCS is proportional to 
\begin{eqnarray}	\no
	\bra{\chi^{\sf sc}_1} \chi^{\sf sc}_0\rangle	
	= i e^{i \bkappa_0 \dot (\br_0 -\br_1)} (\bc^*_\perp \times \bc_\perp) \dot \bra{\chi^\alpha_1}\hat \bsigma \ket{\chi^\alpha_0}.
\end{eqnarray}
For a maximally entangled state, $\bc$ is real-valued and $\bc^*_\perp \times \bc_\perp$ vanishes identically, thus explaining the observed quantum erasure phenomenon. 

There remains a question of how to detect the neutron after it has undergone interaction with the target. 
In principle, the phase, $\phi$, added by the entangler is determined by the experimental setup. While this phase may be 
present in general, it is ubiquitous in neutron scattering experiments and may vary with 
neutron wavelength in some cases \cite{shen2019unveiling}. In that context, to obviate the need to average the 
DCS over $\phi$ a spin echo technique might be used to remove this phase by placing a disentangler 
\cite{lu2019operator,shen2019unveiling} after the target. This quantum detection strategy is known as spin-echo ({\sf se}) measurement. 

However, the details of the required spin echo apparatus will depend on the target, both because the latter will rotate the 
neutron polarization and because the scattering is inelastic.  The neutron polarization induced by the target, which will 
depend on $\phi$, can be found from an expression identical to Eq. \eqref{Eq:Polarization} with the replacement
\begin{eqnarray}
    \hat \bsigma \!\cdot \! 
    \bQp^\dagger(\bkappa_1) \hat \bsigma \hat \bsigma \ \!\cdot  \! \bQp(\bkappa_2, t) \rightarrow \hat \bsigma \!\cdot \! 
    \bQp^\dagger(\bkappa_1) \hat \bsigma_{\sf se} \hat \bsigma \ \!\cdot  \! \bQp(\bkappa_2, t), \no
\end{eqnarray}
where $\hat \bsigma_{\sf se} = U_\phi^\dagger \hat \bsigma U^{\;}_\phi$
is the unitarily transformed spin operator with $U_\phi= \sum_{\nu=0,1} e^{i (-1)^\nu\phi}\ket{\chi_\nu^\alpha}\!\bra{\chi_\nu^\alpha}$, 
where $\alpha$ is the spin-echo axis. 

Only the components of the neutron polarization produced by the target that are perpendicular to the $\alpha$-axis will be 
modified by the spin-echo disentangler and this has to be taken into account in order to measure the full DCS with the phase 
$\phi$ eliminated. In general, the final neutron polarization measured after the disentangler will be reduced in magnitude, just 
as it is for conventional spin echo measurements, and the variation of the polarization with entanglement length $\xi$ will 
contain information about the electron spin correlations in the target, including their state of entanglement. Calculations of 
these effects will be the subject of a future communication.

\ignore{
\begin{eqnarray}    \no
     && {\mathbf P'_{\sf se}}  \times \frac{d^2 \sigma}{d \Omega \, d E_{\bk'}}
    =   \frac{ k' r_0^2 }{8 \pi^3 \hbar I} 
    \int d k_1 \, k_1^3 d \Lambda^*_{\bk_1} d \Lambda_{\bk_2} 
     \sum_\lambda p_\lambda  \\
    && \int_{-\infty}^\infty \! dt e^{ -i  \omega t }
    {\sf Tr} \big[ \hat\rho^\alpha_{\bk_1,\bk_2} \bra{\lambda}  \hat \bsigma \!\cdot \! 
    \bQp^\dagger(\bkappa_1) \hat \bsigma_{\sf se} \hat \bsigma \ \!\cdot  \! \bQp(\bkappa_2, t)\ket{\lambda}  \big ], 
    \no 
\end{eqnarray}
}

\begin{table*}[htb]
\begin{centering}
\renewcommand{\arraystretch}{1.3}
\begin{tabular}{|c|c|c|c|c|c|}
    \hline
     Neutron Probe & Entanglement Length  & Wave packet Size & Target State & $\frac{d^2 \sigma}{d \Omega d E_{\bk'}}\big|_{\tau \to \tau'} $ & Comments\\
    \hline
    \hline
    Unentangled & $\xi=0$ & $\D \to \infty$ & Any & Two-slit pattern & 
    \\
    \hline
    Entangled & $\xi>0$ & $\D \to \infty$ & Thermal or Entangled & Two slit pattern & DCS $\xi$-independent
    \\
    \hline
    Entangled & $\xi>0$ & $\D \to \infty$ & Unentangled & Two-slit pattern & DCS $\xi$-dependent
    \\
    \hline
     Any & Any & $\D > \xi, |\bd_\perp|$ & Any & Two-slit pattern & Same as for $\D \to \infty$
    \\
    \hline
    Entangled & $\xi>0$ & $\D < \xi, |\bd_\perp|$ & Thermal  & No interference pattern &  Isotropic 
    \\
    \hline
    Entangled & $\xi>0$ & $\D <  \xi, |\bd_\perp|$ & Entangled & No interference pattern & $ \frac{d^2 \sigma}{d \Omega d E_{\bk'}}\big|_{\tau \to \tau'}      \propto  1 -|\tilde \bkappa_0 \dot \bc|^2$
    \\
    \hline
    Entangled & $\xi>0$ & $\D <  \xi, |\bd_\perp|$ & Partially entangled & Intermediate & 
        \\
    \hline
    Entangled & $\xi>0$ & $\D < \xi, |\bd_\perp|$ & Unentangled & Two-slit pattern & DCS $\xi$-dependent
    \\
    \hline
\end{tabular}
\end{centering}
    \caption{ Summary of main results illustrating the DCSs, $\frac{d^2 \sigma}{d \Omega d E_{\bk'}}\big|_{\tau \to \tau'}$,  
    resulting from a neutron beam (with Hilbert space ${\cal H}_{\sf probe}={\cal H}_{\sf path}\otimes {\cal H}_{\sf spin}$) scattered from a motionless spin dimer target.  
Different scenarios emerge depending on the relation between the relevant scales of the problem, with  $\xi$  the entanglement length, 
$|\bd_\perp|$ the dimer size,  $\tilde \kappa_0=(\bk_0-\bk')/|\bk_0-\bk'|$, and $\bc$ characterizing the pure target state as defined in Section~\ref{sec:dimer}.}
    \label{table:2}
\end{table*}

\section{Outlook}
\label{sec:Outlook}
%

We formalized a scattering theory for an incident probe whose quantum state is prepared in either a
mode- or multiparticle-entangled fashion. The key idea is to control the intrinsic entanglement among the subsystems of the 
probe, such as its spin and pathways, to learn about the entanglement present in the target. Exploiting 
such control involves several adjustable length and energy scales that compete with those of the target. 
This competition may generate amplification or erasure of interference patterns that betray information about the 
target's entanglement. Together with interferometric methods of quantum detection of the scattered wave, such as 
measurement in a spin-echo mode, entangled probes promise to become a powerful tool for future investigations. 
Quantum imaging techniques \cite{Gatti,Abouraddy,Ono, ReviewQI1,ReviewQI2}
using light exploit the quantum nature of the probe, including entanglement and squeezing, to achieve enhanced 
precision measurement \cite{KOS} and sensing, for instance. These interferometric techniques often treat the interaction 
between the probe and the target in a semiclassical manner. By contrast, the quantum entangled-probe scattering studied in this
paper represents a fully quantum-mechanical treatment of the interaction allowing, for example, the determination of 
spatio-temporal information about correlations in the target's elementary constituents.

In an effort to describe and manipulate the internal degrees of freedom of the probe, one is forced to depart 
from a plane wave description and consider a full-fledged entangled wave packet formulation wherein the transverse 
coherence of the probe is an adjustable variable. When applied to magnetic 
scattering of neutrons, our framework generalizes van Hove's theory. To gain an intuition for the effect of this 
technique, we analyzed the particular 
case of magnetic scattering of a neutron by a spin-dimer target state. 
We find an enhancement in the differential cross section (DCS) when in the regime where the transverse width of the 
incident wave packets is smaller than the entanglement length (i.e., the separation between packets) and the entanglement 
length is tuned to match a magnetic correlation length of the target.
Remarkably, a maximally entangled target state does not show the typical Young-like interference pattern 
that is present for a non-maximally entangled dimer.
This finding can be interpreted as the quantum erasing by a {\it quantum-entangled double-slit} of the interference pattern 
expected from an un-entangled or classical double-slit.  
The reason behind such an interesting quantum erasing effect is traced to the effect that, whenever the target 
state is maximally entangled, the 
orthogonality between incident and scattered neutron spin states corresponding to each path is preserved. 
If this is not the case then there is always some interference between the paths' contributions to the DCS. 

We summarize our main results in Table ~\ref{table:2}. While the most spectacular effects we have found relate to 
the situation in which the wave packet size $\Delta$ is smaller than the probe entanglement length $\xi$, some effects, such as the fact that the 
DCS does not depend on $\xi$ for maximally entangled target states, also persist 
when the intrinsic coherence length of the neutron  (related to the wave packet size in our calculations)  is larger than $\xi$. 
Entanglement lengths up to about 25 $\mu$m have been achieved \cite{deHaan2007} but the inherent coherence length of neutrons  
has been measured to be larger than this value \cite{treimer2006,Majkrzak:2019yke} in several experiments, implying that only the first 
four rows of the table are immediately accessible experimentally. While it is not yet clear to what extent our results are generally 
applicable to other entangled systems, we hope that this theory and future experiments that it informs may shed light on complex 
phases exhibited by novel materials such as multiferroics, unconventional superconductors, quantum spin liquids, and frustrated 
magnets.

\newpage

\appendix

\section{Calculation of Transition Rate}
\label{appendix:born-calculation}

Starting from the transition amplitude measured away from the forward propagation, 	
\begin{eqnarray}
	\langle \psi' | U_I(t, - \infty ) | \psi \rangle \approx - \frac{i}{\hbar} \sum_\bk \tilde{T}_{\psi'\psi_\bk} \tilde{g}(\bk) \frac{e^{i \omega(\psi',\psi_\bk)t + \epsilon t}}{i \omega(\psi',\psi_\bk) + \epsilon},  \nonumber 
\end{eqnarray} 
we square this quantity to obtain the probability of specific transition $\psi \to \psi'$ at time $t$ 
\begin{eqnarray}
	&&|\bra{\psi'} U_I(t, -\infty) \ket{\psi}|^2 = \label{eq:amp2} \\
	&& \hspace{1cm} \frac{1}{\hbar^2} \sum_{\bk_1, \bk_2} \tilde{T}_{\psi' \psi_{\bk_1}}^* \tilde{T}_{\psi' \psi_{\bk_2}} \tilde{g}^*(\bk_1) \tilde{g}(\bk_2) \no \\
	&& \hspace{1.5cm} \times \frac{e^{i (\omega(k_1) - \omega( k_2 )) t + 2 \epsilon t}}{(\omega(\psi', \psi_{\bk_1}) +i \epsilon) (\omega(\psi', \psi_{\bk_2}) -i \epsilon)}.  \no
\end{eqnarray}
These frequencies are defined in Section~\ref{sec:epst}. The denominator may be written as
\begin{eqnarray}
	&& \frac{1}{(\omega(\psi', \psi_{\bk_1}) +i \epsilon) (\omega(\psi', \psi_{\bk_2}) -i \epsilon)} = \\
	&& \frac{1}{(\omega(k_1) \!-\! \omega( k_2 )) \!-\! 2i\epsilon} \!\! \left(\! \frac{1}{\omega(\psi' , \psi_{\bk_1} ) + i \epsilon} \!-\! \frac{1}{\omega(\psi' , \psi_{\bk_2} ) - i \epsilon } \!\right). \no 
\end{eqnarray}
With this substitution, we differentiate Eq.~\eqref{eq:amp2} with respect to time in order to express the transition rate of $\psi \to \psi'$. 
This quantity, once $\epsilon \to 0^+$ is taken, we define in the text as 
\begin{eqnarray} 
	&& W_{\psi \to \psi'}(t) \equiv \lim_{\epsilon \to 0^+} \frac{d}{dt} | \bra{\psi'} U_I(t, -\infty) \ket{\psi} |^2 \no \\
	&& = \lim_{\epsilon \to 0^+} \frac{i}{\hbar^2} \!\!\! \sum_{\bk_1, \bk_2} \!\! \tilde{T}_{\psi' \psi_{\bk_1}}^* \tilde{T}_{\psi' \psi_{\bk_2}} \tilde{g}^*(\bk_1) \tilde{g}(\bk_2) e^{i (\omega(k_1) - \omega( k_2 ))t + 2 \epsilon t}  \no \\
	&& \hspace{1.5cm} \times \left( \frac{1}{\omega(\psi' , \psi_{\bk_1} ) + i \epsilon} - \frac{1}{\omega(\psi' , \psi_{\bk_2} ) - i \epsilon } \right). 
	\label{freqdenom}
\end{eqnarray}
This limit may be resolved by using 
\begin{eqnarray}
	\lim_{\epsilon \to 0^+} \frac{1}{u \mp i \epsilon} &=& \pm i \pi \delta(u) + \mathcal{P} \left( \frac{1}{u} \right) ,
\end{eqnarray}
in which $\mathcal{P}$ denotes the Cauchy principal value (PV). The parentheses in Eq. \eqref{freqdenom}  then become
\begin{eqnarray}
	&& -i \pi \big[ \delta(\omega(\psi', \psi_{\bk_1})) + \delta (\omega(\psi', \psi_{\bk_2}))\big ] \no \\
	&& \hspace{2cm} + \mathcal{P}\left(\frac{1}{\omega(\psi', \psi_{\bk_1})}\right) - \mathcal{P} \left(\frac{1}{\omega(\psi', \psi_{\bk_2} )} \right).  \no 
\end{eqnarray}
The difference of PVs will cancel, as described below, so we omit them from now on. 

In the wave-packet context, and in contrast to the plane-wave context, the time dependence of the transition rate is to be expected. As the wave packet evolves with time from $-\infty$ to $t$, the total probability of the specific transition $\psi \to \psi'$ to occur is given by $\int_{- \infty}^t dt' W_{\psi \to \psi'}(t')$. Because we want the cross section per final-state energy, we multiply $W_{\psi \to \psi'}$ by the density of states $\rho(E_{\bk'}) = \frac{mk'}{\hbar^2} \left( \frac{L}{2\pi} \right)^3 d \Omega_{\bk'}$. Performing this evaluation, the probability of transition from $\psi$ to states surrounding $\psi'$ after a long time $t$ is
\begin{eqnarray}
	&& \rho(E_{\bk'}) \lim_{t \to \infty} \int_{-\infty}^t dt' W_{\psi \to \psi'}(t') \\
	&& = \frac{2 \pi^2 }{\hbar^2} \rho(E_{\bk'}) \sum_{\bk_1, \bk_2} \tilde{g}^*(\bk_1) \tilde{g}(\bk_2) \tilde{T}^*_{\psi' \psi_{\bk_1}} T_{\psi' \psi_{\bk_2}} \no \\
	&& \hspace{0.5cm} \times \delta(\omega(k_1) - \omega( k_2 )) \big [\delta(\omega(\psi', \psi_{\bk_1})) + \delta(\omega(\psi', \psi_{\bk_2})) \big] \no \\
	&&= \frac{4\pi^2  }{\hbar} \rho(E_{\bk'}) \sum_{\bk_1, \bk_2} \tilde{g}^*(\bk_1) \tilde{g}(\bk_2) \tilde{T}^*_{\psi' \psi_{\bk_1}} T_{\psi' \psi_{\bk_2}} \no \\
	&& \hspace{2cm} \times \delta(\omega(k_1) - \omega( k_2 )) \delta(\hbar \omega + E_\lambda - E_{\lambda'} ) \no ,
\end{eqnarray}
where we have used $E_{\bk_1}=E_{\bk_2}$ from the first delta function. This same condition  is the reason for
cancellation of the PVs, as it fixes $\omega(\psi', \psi_{\bk_1}) = \omega(\psi', \psi_{\bk_2})$. The time integration used $\int_{-\infty}^\infty dt' e^{iut'}
= 2 \pi \delta(u)$ for a frequency $u$. The energy transferred $\hbar \omega = E_{\bk_1}- E_{\bk'}$ is equal to the energy acquired by the target, $E_{\lambda'} - E_{\lambda}$. 
Thus Eq.~\eqref{eq:transitionprobability} has been recovered, as well as the subsequent form of the energy-conserving delta function in Eq.~\eqref{eq:main result}.

\section{Scattering of Multipartite Entangled Probe}
\label{appendix:multiparticle}
 
\subsection{Incoming particle-entangled state}
The main manuscript deals primarily with a single-particle probe prepared in a mode-entangled state. This is distinct from a setup involving probe states with multiple entangled particles: the latter may lead to a higher-order correlation function due to the additional instances of single-particle interactions with the target. There are several quantitative adjustments which effect this difference in the case of multiparticle scattering. 

First, the Hilbert space of the system is extended to accommodate the momentum and spin states of the second particle.  Indistinguishable particles impose an additional superselection rule on the state space, that is, the exchange statistics symmetry. Here, we only focus on fermionic probes. Indexing the two particle orbitals by $A$ and $B$, we notate a basis for the state of two spin-$\frac{1}{2}$ particles of momenta $\bk_A$ and $\bk_B$ and entanglement vector $\bxi$: 
\begin{eqnarray}
    \mathcal{B} &=& \left\{ a^{\bxi}_{\bk_A \bk_B} (\bx_A, \bx_B), b^{\bxi}_{\bk_A \bk_B} (\bx_A, \bx_B), c^{\nu}_{\bk_A \bk_B}(\bx_A, \bx_B)   \right\}, \no
\end{eqnarray}
where $\bx_i = (\br_i, \sigma_i)$ labels the position and spin of the particle $i=A,B$, and $\nu=0,1$. These basis functions obey the fermionic superselection rule, antisymmetry under exchange $\bx_A \leftrightarrow \bx_B$, and are defined by
\begin{eqnarray}
    a^{\bxi}_{\bk_A \bk_B} (\bx_A, \bx_B) &=& \frac{ 1 }{2L^{3} \sqrt{1 + \delta_{\bk_A \bk_B}}} \no \\
    && \hspace{-2.5cm} \times \left[e^{i(\bk_A \cdot \br_A + \bk_B \cdot \br_B )} \sd^- (\bk_A, \sigma_A, \bk_B, \sigma_B, \bxi) + ( \bk_A \leftrightarrow \bk_B) \right], \no \\
    b^{\bxi}_{\bk_A \bk_B} (\bx_A, \bx_B) &=& \frac{ 1 }{2L^{3}} \no \\ 
    && \hspace{-2.5cm} \times \left[ e^{i(\bk_A \cdot \br_A + \bk_B \cdot \br_B) } \sd^+ (\bk_A, \sigma_A, \bk_B, \sigma_B, \bxi) - ( \bk_A \leftrightarrow \bk_B) \right], \no \\
    c^{\nu}_{\bk_A \bk_B} (\bx_A, \bx_B) &=& \frac{ 1 }{\sqrt{2}L^{3}} \no \\
    &&\hspace{-2.5cm} \times \left[ e^{i ( \bk_A \cdot \br_A + \bk_B \cdot \br_B )} \chi^\alpha_\nu (\sigma_A) \chi^\alpha_{\nu} (\sigma_B) - ( \bk_A \leftrightarrow \bk_B) \right] ,\no
\end{eqnarray}
with 
\begin{eqnarray}
    \sd^\pm (\bk_A, \sigma_A, \bk_B, \sigma_B, \bxi) &=& \no \\
    &&\hspace{-1.5cm} e^{- \frac{i}{2} (\bk_A - \bk_B) \cdot \bxi} \chi^\alpha_0 (\sigma_A) \chi^\alpha_1 (\sigma_B) \no \\
    && \hspace{-0.5cm} \pm e^{\frac{i}{2} (\bk_A - \bk_B) \cdot \bxi} \chi^\alpha_1 (\sigma_A) \chi^\alpha_0 (\sigma_B). \no
\end{eqnarray}
It is worth noting that $b^{\bxi}_{\bk \bk} = c^{\nu}_{\bk \bk} = 0$ and so are not included in this basis. $\chi^\alpha_{\nu} (\sigma_i)$ is the spinor of the $i$ particle aligned up ($\nu = 0$) or down ($\nu = 1$) along the $\alpha$ quantization axis. 

To illustrate the effect of particle entanglement we define an example initial state from the basis vectors to be
\begin{eqnarray}
    \bra{\bx_A, \bx_B } \Psi_{\sf in} \rangle &=& \frac{1}{2} \sum_{\bk_A, \bk_B} \tilde{g} (\bk_A) \tilde{g}(\bk_B) a^\bxi_{\bk_A, \bk_B} (\bx_A, \bx_B).  \no 
\end{eqnarray}
$H_{\sf p} = - \frac{\hbar^2}{2m} ( \nabla_{\br_A}^2 + \nabla_{\br_B}^2)$ is the Hamiltonian of the free probes with momenta $\bk_A$ and $\bk_B$. 

\begin{figure}[thb]
    \centering
    \includegraphics[height=0.50\columnwidth]{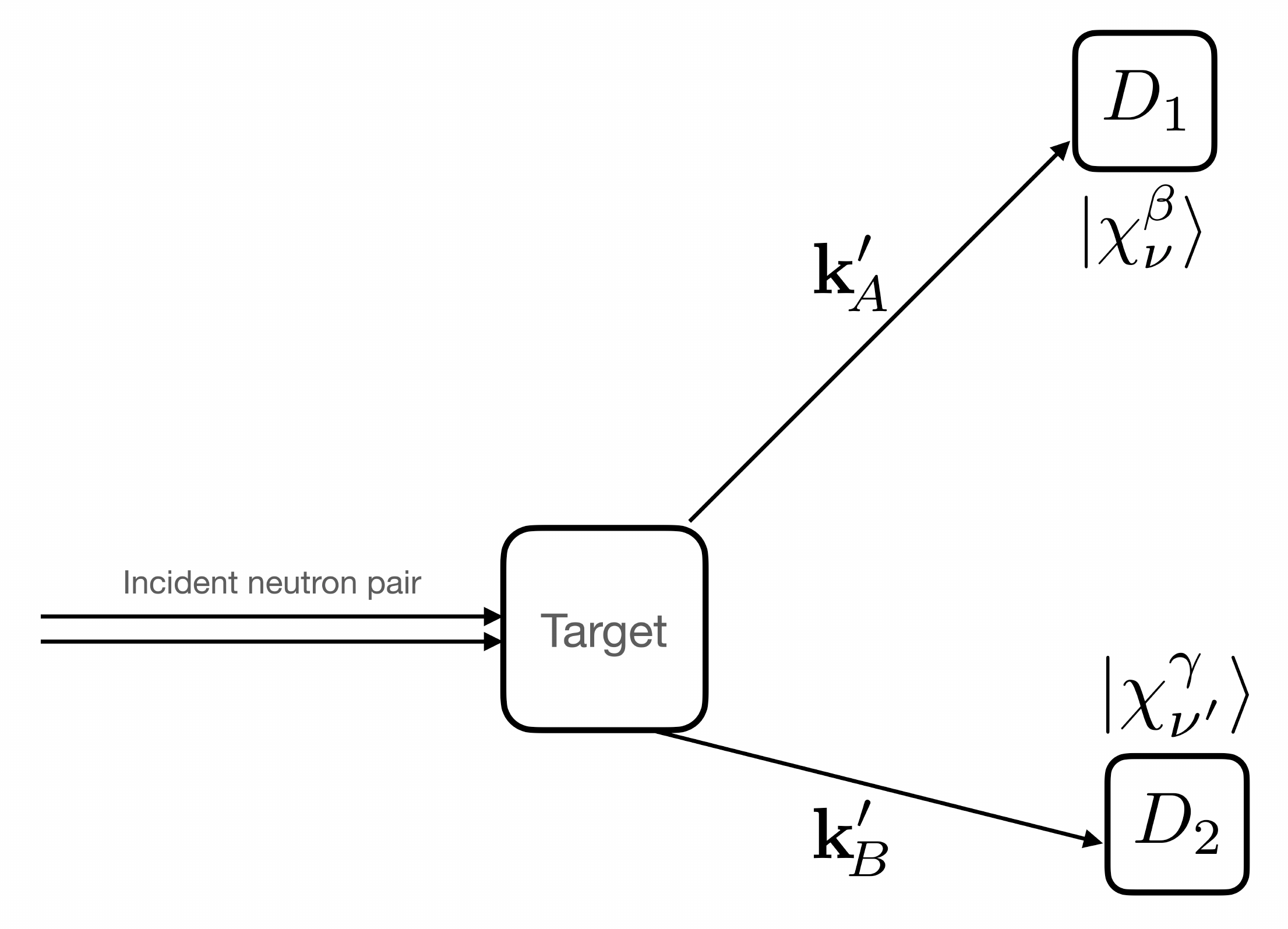} 
    \caption{We have two detectors $D_1$ and $D_2$, spin-resolved along the $\beta$ and $\gamma$ spin quantization axis. Neutrons are counted when neutrons hit the different detectors simultaneously. }
    \label{fig: multipartite detection}
\end{figure}

We next describe the final state which is detectable by the set-up of Fig.\ref{fig: multipartite detection}: two detectors separated in space are polarized to only detect up or down ($\nu, \nu' = 0, 1$) as seen along the axes $\beta$ and $\g$:
\begin{eqnarray}
    \bra{\bx_A, \bx_B} \Psi_{\sf out} \rangle &=& \frac{1}{ L^3 \sqrt 2} \big(e^{i (\bk'_{\! A} \cdot \br_A + \bk'_{\! B} \cdot \br_B)}  \chi^{\beta}_\nu (\sigma_A) \chi^{\g}_{\nu'} (\sigma_B) \no \\
    && -e^{i (\bk'_B \cdot \br_A + \bk'_A \cdot \br_B)}  
    \chi^{\g}_{\nu'} (\sigma_A) \chi^{\beta}_{\nu} (\sigma_B)
    \big). \label{eq:outstate}
\end{eqnarray}
A spinor written in the basis along the axis $\beta = (\theta', \phi')$ is connected to the $\alpha = (\theta, \phi)$ basis by the rotation
\begin{eqnarray}
    \chi^{\beta}_{\nu} &=& \mathcal{R}^{\beta}_{\nu 0} \chi^{\alpha}_0 + \mathcal{R}^{\beta}_{\nu 1} \chi^\alpha_1
    ,\no 
\end{eqnarray}
with rotation matrix 
\begin{eqnarray}
    \mathcal{R}^{\beta} = \begin{pmatrix}
        c_{\theta'} c_{\theta} + e^{i (\phi'-\phi)} s_{\theta'} s_{\theta} && - c_{\theta'} s_{\theta} + e^{i (\phi'-\phi)} s_{\theta'} c_{\theta} \\
        - s_{\theta'} c_{\theta} + e^{i(\phi'-\phi)} c_{\theta'} s_{\theta} && s_{\theta'} s_{\theta} + e^{i (\phi'-\phi)} c_{\theta'} c_{\theta} \no
    \end{pmatrix}
\end{eqnarray}
and $\chi^{\a}_{0}=
\begin{pmatrix}
c_{\theta}
\\
e^{i \phi } s_{\theta}      
\end{pmatrix}$, $\chi^{\a}_{1}=
\begin{pmatrix}
      -s_{\theta}
      \\
      e^{i \phi } c_{\theta}
\end{pmatrix}
$, $s_{\theta} \equiv \sin {\frac{\theta}{2}}$ and $c_{\theta} \equiv \cos {\frac{\theta}{2}}$. Substituting these expressions in Eq.~\eqref{eq:outstate} we find that this {\sf out} state can be written in terms of $\bxi=0$ basis functions as 
\begin{eqnarray}
    \bra{\bx_A, \bx_B} \Psi_{\sf out} \rangle &=& \sum_{\mu=0,1} \mathcal{R}^{\beta}_{\nu \mu} \mathcal{R}^{\g}_{\nu' \mu} c^\mu_{\bk_A \bk_B} (\bx_A, \bx_B) \no \\ 
    && \hspace{-1cm} + \frac{1}{\sqrt{2}} \Big( \mathcal{R}^{\beta}_{\nu 0} \mathcal{R}^{\g}_{\nu' 1} \big( a^0_{\bk_A \bk_B} + b^0_{\bk_A \bk_B} \big) \no \\
    && \hspace{1.5cm} + \mathcal{R}^{\beta}_{\nu 1} \mathcal{R}^{\g}_{\nu' 0} \big( b^0_{\bk_A \bk_B} - a^0_{\bk_A \bk_B} \big) \Big). \no
\end{eqnarray}

The interaction potential is now extended to include a symmetric interaction $\hat V$ between the two particles. Additionally, we take it to obey the locality condition
\begin{eqnarray*}
    \bra{\br_A', \br_B'} \hat{V} \ket{\br_A, \br_B} = \delta(\br_A' - \br_A) \delta(\br_B' - \br_B) \hat{V}(\bx_A, \bx_B)
\end{eqnarray*}
with $\hat{V}(\bx_A, \bx_B) = \hat{V} (\bx_B, \bx_A)$, i.e., a symmetric potential. Calculating the potential matrix element between our initial and final states, 
\begin{widetext}
\begin{eqnarray}
    \bra{\Psi_{\sf out} } \hat{V} \ket{\Psi_{\sf in} } &=& \int d\br_A d\br_B \bra{\Psi_{\sf out} } \br_A, \br_B \rangle \bra{ \br_A, \br_B } \hat{V} \ket{\br_A, \br_B} \bra{\br_A, \br_B} \Psi_{\sf in} \rangle 
    \no \\ 
    &=& \frac{1}{4 L^6} \sum_{\bk_A \bk_B} \frac{\tilde{g} (\bk_A) \tilde{g} (\bk_B)}{2^{\frac{1+\delta_{\bk_A \bk_B}}{2}}} \int d\br_A d\br_B \bra{\cdots} \hat{V} \ket{\cdots} 
    \no     \\
    &&\times     \big[  e^{-i ( \bk_A' \cdot \br_A + \bk_B' \cdot \br_B) } \chi^{\beta}_\nu (\sigma_A)^\dagger \chi^{\gamma}_{\nu'}(\sigma_B)^\dagger - e^{-i(\bk_B' \cdot \br_A + \bk_A' \cdot \br_B)} \chi^{\gamma}_{\nu'} (\sigma_A)^\dagger \chi^{\beta}_{\nu} (\sigma_B)^\dagger \big]  
    \no \\
    && \times \big[  e^{i(\bk_A \cdot \br_A + \bk_B \cdot \br_B) } \big( e^{-\frac{i}{2} (\bk_A - \bk_B) \cdot \bxi} \chi^\alpha_0 (\sigma_A) \chi^\alpha_1 (\sigma_B) - e^{\frac{i}{2} (\bk_A - \bk_B) \cdot \bxi} \chi^\alpha_1(\sigma_A) \chi^\alpha_0 (\sigma_B) \big)  
     + ( \bk_A \leftrightarrow \bk_B) \big] 
    \no \\
    &=& \frac{1}{ 4 L^6 } \sum_{\bk_A \bk_B} \frac{\tilde{g} (\bk_A) \tilde{g} (\bk_B)}{2^{\frac{1+\delta_{\bk_A \bk_B}}{2}}} \int d\br_A d\br_B \bra{\cdots} \hat{V} \ket{\cdots} 
    \no \\
    && \times \bigg[  e^{i ( (\bk_A - \bk_A') \cdot \br_A + (\bk_B - \bk_B' ) \cdot \br_B ) } e^{- \frac{i}{2} (\bk_A - \bk_B) \cdot \bxi} \left( \chi^{\beta \dagger}_\nu \chi^\alpha_0 (\sigma_A) \right) \left( \chi^{\gamma \dagger}_{\nu'} \chi^{\alpha}_1(\sigma_B) \right) \no \\
    && \hspace{2cm} + e^{i (( \bk_A - \bk_B') \cdot \br_A + (\bk_B - \bk_A') \cdot \br_B )}e^{\frac{i}{2} (\bk_A - \bk_B) \cdot \bxi} \left( \chi^{\gamma \dagger}_{\nu'} \chi^\alpha_1 (\sigma_A) \right) \left( \chi^{\beta \dagger}_\nu \chi^\alpha_0 (\sigma_B) \right) \no \\
    && \hspace{2cm} - e^{i((\bk_A - \bk_B')\cdot \br_A + (\bk_B - \bk_A') \cdot \br_B) } e^{-\frac{i}{2} (\bk_A - \bk_B) \cdot \bxi} \left( \chi^{\gamma \dagger}_{\nu'} \chi^\alpha_0 (\sigma_A) \right) \left( \chi^{\beta \dagger}_\nu \chi^{\alpha}_1 (\sigma_B) \right) \no \\
    && \hspace{2cm} - e^{i ((\bk_A - \bk_A') \cdot \br_A + (\bk_B - \bk_B') \cdot \br_B )} e^{\frac{i}{2} (\bk_A - \bk_B) \cdot \bxi} \left( \chi^{\beta \dagger}_\nu \chi^{\alpha}_1( \sigma_A) \right) \left( \chi^{\gamma \dagger}_{\nu'} \chi^\alpha_0 (\sigma_B) \right) 
     + ( \bk_A \leftrightarrow \bk_B) \bigg] 
    \no \\
    &\equiv& \frac{1}{4 L^6} \sum_{\bk_A \bk_B} \frac{\tilde{g} (\bk_A) \tilde{g} ( \bk_B )}{2^{\frac{1+\delta_{\bk_A \bk_B}}{2}}} V^{\nu \nu'}_{{\sf out}, \bk_A \bk_B}. \no 
\end{eqnarray}
\end{widetext}
The spin components of expressions such as this will be determined by the initial and final configurations, averaged and summed over as usual in the cross-section. 
\subsection{Final two-fermions cross-section}
The amplitude for transition from the initially prepared probe to a final basis state, again measuring away from the forward direction, is
\begin{eqnarray}
    && \bra{\Psi_{\sf out}} U_I(t, -\infty ) \ket{\Psi_{\sf in}} \approx \no \\ 
    && \hspace{0cm} -\frac{1}{4L^6} \frac{i}{\hbar} \sum_{\bk_A, \bk_B} \frac{\tilde{g}(\bk_A) \tilde{g} (\bk_B) }{2^{\frac{1+\delta_{\bk_A \bk_B}}{2}}} V^{\nu\nu'}_{{\sf out},\bk_A \bk_B} \int_{- \infty}^t dt \, e^{i \omega_{\Psi_{\sf out},\bk_A \bk_B} t + \epsilon t} \no
\end{eqnarray}
with 
\begin{eqnarray}
    \hbar \omega_{\Psi', (\bk_A, \bk_B)} = (E_{\lambda'} - E_\lambda) + \frac{\hbar^2}{2m} (\bk_A'^2 + \bk_B'^2 - \bk_A^2 - \bk_B^2). \no
\end{eqnarray}
Following the same procedure as in Appendix \ref{appendix:born-calculation} from here, the probability of transition from $\Psi_{\sf in}$ to states surrounding $\Psi_{\sf out}$ per final state energies $E_{\bk_A'}$ \emph{and} $E_{\bk_B'}$ after a long time $t$ is 
\begin{eqnarray}
    &&\rho(E_{\bk_A'}) \rho(E_{\bk_B'}) \lim_{t \to \infty} \int_{-\infty }^t dt' W_{\Psi_{\sf in} \to \Psi_{\sf out}}(t') \no 
\end{eqnarray}
with now 
\begin{eqnarray}
	W_{\Psi_{\sf in} \to \Psi_{\sf out}} \equiv \lim_{\epsilon \to 0^+} \frac{d}{dt} | \bra{\Psi_{\sf out} } U_I(t, -\infty) \ket{\Psi_{\sf in} } |^2 \no
\end{eqnarray}
and the density of states being for a pair of plane waves of individual energies
\begin{eqnarray}
    \rho(E_{\bk_A'}) \rho(E_{\bk_B'}) &=& \frac{m^2 k_A' k_B'}{\hbar^4} \left( \frac{L}{2\pi} \right)^6 d\Omega_{\bk_A'} d\Omega_{\bk_B'}. \no
\end{eqnarray}
The flux, now, is the combined contribution to the time-integrated flux by the two probes $A$ and $B$ averaged over a characteristic area: 
$I_A + I_B = \int_{-\infty}^{\infty} dt \, (\overline{\jmath}_A + \overline{\jmath}_B) $. Taking also the sum over final target and spin states and 
averaging over initial target states, the two-particle cross section is now
\begin{widetext}
\begin{eqnarray}
    &&\frac{d^4 \sigma}{dE_{k_A'} dE_{k_B'} d\Omega_{\bk_A'} d\Omega_{\bk_B'} }  
    \no \\
    &=& \tilde{\mathcal{C}}  \sum_{\substack{\lambda, \lambda' \\ \nu, \nu'}} p_\lambda \int d\Lambda_{A1} d\Lambda_{B1} d\Lambda_{A2}^* d\Lambda_{B2}^* 
    \delta( \hbar \omega_{\Psi_{\sf out}, (\bk_{A1}, \bk_{B1})} ) \delta( \hbar \omega_{(\bk_{A1}, \bk_{B1}),(\bk_{A2},\bk_{B2} ) } ) V^{\nu \nu' *}_{{\sf out}, \bk_{A2} \bk_{B2}} V^{\nu \nu'}_{{\sf out}, \bk_{A1}\bk_{B1}} \nonumber
     \\
    &=& \tilde{\mathcal{C}} \sum_{\substack{\lambda, \lambda' \\
    \nu, \nu'}} p_\lambda \int d\Lambda_{A1} d\Lambda_{B1} d\Lambda_{A2}^* d\Lambda_{B2}^* \delta( \hbar \omega_{\Psi_{\sf out}, (\bk_{A1}, \bk_{B1})} ) \delta( \hbar \omega_{(\bk_{A1}, \bk_{B2}),(\bk_{A2},\bk_{B2} ) } ) \no \\
    && \hspace{6cm} \times \int d\br_{A1} d\br_{B1} d\br_{A2} d\br_{B2} G_{\lambda, \lambda', \bxi}^{\nu, \nu'} (\bx_{A1}, \bx_{B1}, \bx_{A2}, \bx_{B2}) 
\end{eqnarray}
\end{widetext}
with 
$ \tilde{\mathcal{C}}=\frac{m^2 k_A' k_B'}{16 (2\pi)^4 \hbar^2 (I_A+I_B) }$,
$d\Lambda_{i} \equiv g(\bk_i) d \bk_i$ (which differs from the main text in that it includes also the magnitude integration) and the four-point spatial correlation function given by
\begin{widetext}
\begin{eqnarray}
    G_{\lambda, \lambda', \bxi}^{\nu, \nu'} (\bx_{A1}, \bx_{B1}, \bx_{A2}, \bx_{B2})&=&
    \no     \\
    && \hspace*{-2cm}=  a^{0}_{\bk_A', \bk_B'} (\bx_{A2}, \bx_{B2})^\dagger a^0_{\bk_A', \bk_B'} (\bx_{A1}, \bx_{B1})  
      a^\bxi_{\bk_{A2},
      \bk_{B2}} (\bx_{A2}, \bx_{B2} )^\dagger a^\bxi_{\bk_{A1}, \bk_{B1}}(\bx_{A1}, \bx_{B1})  
     \no    \\
     && \hspace*{-1cm}\times V^{\nu \nu' *}_{{\sf out}, \bk_{A2} \bk_{B2}} (\bx_{A2}, \bx_{B2}) V^{\nu \nu'}_{{\sf out}, \bk_{A1}\bk_{B1}}(\bx_{A1}, \bx_{B1}). 
\end{eqnarray}
\end{widetext}

\bigskip
\begin{acknowledgments}
 We are grateful to Mike Snow for illuminating discussions. We would also thank the ``IU neutron team'', D. V. Baxter, E. Dees, S. Kuhn, S. McKay, and J. Shen for constructive and stimulating exchanges. The IU Quantum Science and Engineering Center is supported by the Office of the IU Bloomington Vice Provost for Research through its Emerging Areas of Research program. 
\end{acknowledgments}

\bibliographystyle{apsrev4-1}
\bibliography{BIB}

\begin{thebibliography}{23}%
\makeatletter
\providecommand \@ifxundefined [1]{%
 \@ifx{#1\undefined}
}%
\providecommand \@ifnum [1]{%
 \ifnum #1\expandafter \@firstoftwo
 \else \expandafter \@secondoftwo
 \fi
}%
\providecommand \@ifx [1]{%
 \ifx #1\expandafter \@firstoftwo
 \else \expandafter \@secondoftwo
 \fi
}%
\providecommand \natexlab [1]{#1}%
\providecommand \enquote  [1]{``#1''}%
\providecommand \bibnamefont  [1]{#1}%
\providecommand \bibfnamefont [1]{#1}%
\providecommand \citenamefont [1]{#1}%
\providecommand \href@noop [0]{\@secondoftwo}%
\providecommand \href [0]{\begingroup \@sanitize@url \@href}%
\providecommand \@href[1]{\@@startlink{#1}\@@href}%
\providecommand \@@href[1]{\endgroup#1\@@endlink}%
\providecommand \@sanitize@url [0]{\catcode `\\12\catcode `\$12\catcode
  `\&12\catcode `\#12\catcode `\^12\catcode `\_12\catcode `\%12\relax}%
\providecommand \@@startlink[1]{}%
\providecommand \@@endlink[0]{}%
\providecommand \url  [0]{\begingroup\@sanitize@url \@url }%
\providecommand \@url [1]{\endgroup\@href {#1}{\urlprefix }}%
\providecommand \urlprefix  [0]{URL }%
\providecommand \Eprint [0]{\href }%
\providecommand \doibase [0]{http://dx.doi.org/}%
\providecommand \selectlanguage [0]{\@gobble}%
\providecommand \bibinfo  [0]{\@secondoftwo}%
\providecommand \bibfield  [0]{\@secondoftwo}%
\providecommand \translation [1]{[#1]}%
\providecommand \BibitemOpen [0]{}%
\providecommand \bibitemStop [0]{}%
\providecommand \bibitemNoStop [0]{.\EOS\space}%
\providecommand \EOS [0]{\spacefactor3000\relax}%
\providecommand \BibitemShut  [1]{\csname bibitem#1\endcsname}%
\let\auto@bib@innerbib\@empty
\bibitem [{\citenamefont {Shen}\ \emph {et~al.}(2019)\citenamefont {Shen},
  \citenamefont {Kuhn}, \citenamefont {Dalgliesh}, \citenamefont {de~Haan},
  \citenamefont {Geerits}, \citenamefont {Irfan}, \citenamefont {Li},
  \citenamefont {Lu}, \citenamefont {Parnell}, \citenamefont {Plomp},
  \citenamefont {van Well}, \citenamefont {Washington}, \citenamefont {Baxter},
  \citenamefont {Ortiz}, \citenamefont {Snow},\ and\ \citenamefont
  {Pynn}}]{shen2019unveiling}%
  \BibitemOpen
  \bibfield  {author} {\bibinfo {author} {\bibfnamefont {J.}~\bibnamefont
  {Shen}}, \bibinfo {author} {\bibfnamefont {S.~J.}\ \bibnamefont {Kuhn}},
  \bibinfo {author} {\bibfnamefont {R.~M.}\ \bibnamefont {Dalgliesh}}, \bibinfo
  {author} {\bibfnamefont {V.~O.}\ \bibnamefont {de~Haan}}, \bibinfo {author}
  {\bibfnamefont {N.}~\bibnamefont {Geerits}}, \bibinfo {author} {\bibfnamefont
  {A.~A.~M.}\ \bibnamefont {Irfan}}, \bibinfo {author} {\bibfnamefont
  {F.}~\bibnamefont {Li}}, \bibinfo {author} {\bibfnamefont {S.}~\bibnamefont
  {Lu}}, \bibinfo {author} {\bibfnamefont {S.~R.}\ \bibnamefont {Parnell}},
  \bibinfo {author} {\bibfnamefont {J.}~\bibnamefont {Plomp}}, \bibinfo
  {author} {\bibfnamefont {A.~A.}\ \bibnamefont {van Well}}, \bibinfo {author}
  {\bibfnamefont {A.}~\bibnamefont {Washington}}, \bibinfo {author}
  {\bibfnamefont {D.~V.}\ \bibnamefont {Baxter}}, \bibinfo {author}
  {\bibfnamefont {G.}~\bibnamefont {Ortiz}}, \bibinfo {author} {\bibfnamefont
  {W.~M.}\ \bibnamefont {Snow}}, \ and\ \bibinfo {author} {\bibfnamefont
  {R.}~\bibnamefont {Pynn}},\ }\href@noop {} {\bibfield  {journal} {\bibinfo
  {journal} {Nature Communications}\ }\textbf {\bibinfo {volume} {11}},\
  \bibinfo {pages} {930} (\bibinfo {year} {2019})}\BibitemShut {NoStop}%
\bibitem [{\citenamefont {Van~Hove}(1954)}]{vanhove1954}%
  \BibitemOpen
  \bibfield  {author} {\bibinfo {author} {\bibfnamefont {L.}~\bibnamefont
  {Van~Hove}},\ }\href {\doibase 10.1103/PhysRev.95.249} {\bibfield  {journal}
  {\bibinfo  {journal} {Phys. Rev.}\ }\textbf {\bibinfo {volume} {95}},\
  \bibinfo {pages} {249} (\bibinfo {year} {1954})}\BibitemShut {NoStop}%
\bibitem [{\citenamefont {Lu}\ \emph {et~al.}(2020)\citenamefont {Lu},
  \citenamefont {Irfan}, \citenamefont {Shen}, \citenamefont {Kuhn},
  \citenamefont {Snow}, \citenamefont {Baxter}, \citenamefont {Pynn},\ and\
  \citenamefont {Ortiz}}]{lu2019operator}%
  \BibitemOpen
  \bibfield  {author} {\bibinfo {author} {\bibfnamefont {S.}~\bibnamefont
  {Lu}}, \bibinfo {author} {\bibfnamefont {A.~A.~M.}\ \bibnamefont {Irfan}},
  \bibinfo {author} {\bibfnamefont {J.}~\bibnamefont {Shen}}, \bibinfo {author}
  {\bibfnamefont {S.~J.}\ \bibnamefont {Kuhn}}, \bibinfo {author}
  {\bibfnamefont {W.~M.}\ \bibnamefont {Snow}}, \bibinfo {author}
  {\bibfnamefont {D.~V.}\ \bibnamefont {Baxter}}, \bibinfo {author}
  {\bibfnamefont {R.}~\bibnamefont {Pynn}}, \ and\ \bibinfo {author}
  {\bibfnamefont {G.}~\bibnamefont {Ortiz}},\ }\href@noop {} {\bibfield
  {journal} {\bibinfo  {journal} {Phys. Rev. A}\ }\textbf {\bibinfo {volume}
  {101}},\ \bibinfo {pages} {042318} (\bibinfo {year} {2020})}\BibitemShut
  {NoStop}%
\bibitem [{\citenamefont {Kuhn}\ \emph {et~al.}(2021)\citenamefont {Kuhn},
  \citenamefont {McKay}, \citenamefont {Shen}, \citenamefont {Geerits},
  \citenamefont {Dalgliesh}, \citenamefont {Dees}, \citenamefont {Irfan},
  \citenamefont {Li}, \citenamefont {Lu}, \citenamefont {Vangelista},
  \citenamefont {Baxter}, \citenamefont {Ortiz}, \citenamefont {Parnell},
  \citenamefont {Snow},\ and\ \citenamefont {Pynn}}]{kuhn2020unveiling}%
  \BibitemOpen
  \bibfield  {author} {\bibinfo {author} {\bibfnamefont {S.~J.}\ \bibnamefont
  {Kuhn}}, \bibinfo {author} {\bibfnamefont {S.}~\bibnamefont {McKay}},
  \bibinfo {author} {\bibfnamefont {J.}~\bibnamefont {Shen}}, \bibinfo {author}
  {\bibfnamefont {N.}~\bibnamefont {Geerits}}, \bibinfo {author} {\bibfnamefont
  {R.~M.}\ \bibnamefont {Dalgliesh}}, \bibinfo {author} {\bibfnamefont
  {E.}~\bibnamefont {Dees}}, \bibinfo {author} {\bibfnamefont {A.~A.~M.}\
  \bibnamefont {Irfan}}, \bibinfo {author} {\bibfnamefont {F.}~\bibnamefont
  {Li}}, \bibinfo {author} {\bibfnamefont {S.}~\bibnamefont {Lu}}, \bibinfo
  {author} {\bibfnamefont {V.}~\bibnamefont {Vangelista}}, \bibinfo {author}
  {\bibfnamefont {D.~V.}\ \bibnamefont {Baxter}}, \bibinfo {author}
  {\bibfnamefont {G.}~\bibnamefont {Ortiz}}, \bibinfo {author} {\bibfnamefont
  {S.~R.}\ \bibnamefont {Parnell}}, \bibinfo {author} {\bibfnamefont {W.~M.}\
  \bibnamefont {Snow}}, \ and\ \bibinfo {author} {\bibfnamefont
  {R.}~\bibnamefont {Pynn}},\ }\href {\doibase
  10.1103/PhysRevResearch.3.023227} {\bibfield  {journal} {\bibinfo  {journal}
  {Phys. Rev. Research}\ }\textbf {\bibinfo {volume} {3}},\ \bibinfo {pages}
  {023227} (\bibinfo {year} {2021})}\BibitemShut {NoStop}%
\bibitem [{\citenamefont {Lovesey}(2003)}]{loveseybook}%
  \BibitemOpen
  \bibfield  {author} {\bibinfo {author} {\bibfnamefont {S.~W.}\ \bibnamefont
  {Lovesey}},\ }\href {https://books.google.com/books?id=xfLvAAAAMAAJ} {\emph
  {\bibinfo {title} {Theory of Neutron Scattering from Condensed Matter}}},\
  \bibinfo {number} {volumes 1 and 2}\ (\bibinfo  {publisher} {Oxford
  University Press, Oxford},\ \bibinfo {year} {2003})\BibitemShut {NoStop}%
\bibitem [{\citenamefont {Sakurai}\ and\ \citenamefont
  {Napolitano}(2021)}]{Sakuraibook}%
  \BibitemOpen
  \bibfield  {author} {\bibinfo {author} {\bibfnamefont {J.~J.}\ \bibnamefont
  {Sakurai}}\ and\ \bibinfo {author} {\bibfnamefont {J.}~\bibnamefont
  {Napolitano}},\ }\href@noop {} {\emph {\bibinfo {title} {{Modern quantum
  mechanics; 3rd ed.}}}}\ (\bibinfo  {publisher} {Cambridge University Press},\
  \bibinfo {address} {Cambridge},\ \bibinfo {year} {2021})\BibitemShut
  {NoStop}%
\bibitem [{\citenamefont {Newton}(1966)}]{newtonbook}%
  \BibitemOpen
  \bibfield  {author} {\bibinfo {author} {\bibfnamefont {R.~G.}\ \bibnamefont
  {Newton}},\ }\href@noop {} {\emph {\bibinfo {title} {Scattering Theory of
  Waves and Particles}}}\ (\bibinfo  {publisher} {McGraw-Hill, New York},\
  \bibinfo {year} {1966})\BibitemShut {NoStop}%
\bibitem [{\citenamefont {Joachain}(1983)}]{joachain}%
  \BibitemOpen
  \bibfield  {author} {\bibinfo {author} {\bibfnamefont {C.~J.}\ \bibnamefont
  {Joachain}},\ }\href@noop {} {\emph {\bibinfo {title} {Quantum Collision
  Theory}}}\ (\bibinfo  {publisher} {North Holland, Amsterdam},\ \bibinfo
  {year} {1983})\BibitemShut {NoStop}%
\bibitem [{\citenamefont {Barnum}\ \emph {et~al.}(2003)\citenamefont {Barnum},
  \citenamefont {Knill}, \citenamefont {Ortiz},\ and\ \citenamefont
  {Viola}}]{barnum-2003}%
  \BibitemOpen
  \bibfield  {author} {\bibinfo {author} {\bibfnamefont {H.}~\bibnamefont
  {Barnum}}, \bibinfo {author} {\bibfnamefont {E.}~\bibnamefont {Knill}},
  \bibinfo {author} {\bibfnamefont {G.}~\bibnamefont {Ortiz}}, \ and\ \bibinfo
  {author} {\bibfnamefont {L.}~\bibnamefont {Viola}},\ }\href {\doibase
  10.1103/PhysRevA.68.032308} {\bibfield  {journal} {\bibinfo  {journal} {Phys.
  Rev. A}\ }\textbf {\bibinfo {volume} {68}},\ \bibinfo {pages} {032308}
  (\bibinfo {year} {2003})}\BibitemShut {NoStop}%
\bibitem [{\citenamefont {Barnum}\ \emph {et~al.}(2004)\citenamefont {Barnum},
  \citenamefont {Knill}, \citenamefont {Ortiz}, \citenamefont {Somma},\ and\
  \citenamefont {Viola}}]{barnum-2004}%
  \BibitemOpen
  \bibfield  {author} {\bibinfo {author} {\bibfnamefont {H.}~\bibnamefont
  {Barnum}}, \bibinfo {author} {\bibfnamefont {E.}~\bibnamefont {Knill}},
  \bibinfo {author} {\bibfnamefont {G.}~\bibnamefont {Ortiz}}, \bibinfo
  {author} {\bibfnamefont {R.}~\bibnamefont {Somma}}, \ and\ \bibinfo {author}
  {\bibfnamefont {L.}~\bibnamefont {Viola}},\ }\href@noop {} {\bibfield
  {journal} {\bibinfo  {journal} {Phys. Rev. Lett.}\ }\textbf {\bibinfo
  {volume} {92}},\ \bibinfo {pages} {107902} (\bibinfo {year}
  {2004})}\BibitemShut {NoStop}%
\bibitem [{\citenamefont {Schotland}\ \emph {et~al.}(2016)\citenamefont
  {Schotland}, \citenamefont {Caz{\'{e}}},\ and\ \citenamefont
  {Norris}}]{Schotland2016}%
  \BibitemOpen
  \bibfield  {author} {\bibinfo {author} {\bibfnamefont {J.~C.}\ \bibnamefont
  {Schotland}}, \bibinfo {author} {\bibfnamefont {A.}~\bibnamefont
  {Caz{\'{e}}}}, \ and\ \bibinfo {author} {\bibfnamefont {T.~B.}\ \bibnamefont
  {Norris}},\ }\href {\doibase 10.1364/ol.41.000444} {\bibfield  {journal}
  {\bibinfo  {journal} {Optics Letters}\ }\textbf {\bibinfo {volume} {41}},\
  \bibinfo {pages} {444} (\bibinfo {year} {2016})},\ \Eprint
  {http://arxiv.org/abs/1509.07931} {arXiv:1509.07931} \BibitemShut {NoStop}%
\bibitem [{1No()}]{1Note}%
  \BibitemOpen
  \href@noop {} {}\bibinfo {note} {A more general state realized in RF-flipper
  entanglers is \begin{eqnarray} \Phi_\text{in}(\br,t_0) &=&
  \frac{1}{L^\frac{3}{2}}\sum_\bk \tilde g(\bk) e^{i \bk \cdot \br} e^{- i
  \omega(k) t_0} \ \ket{\chi_{\bk\cdot\bxi}}, \no \\ \mbox{ with } \quad
  \ket{\chi_{\bk\cdot\bxi}}&=&\frac{ e^{- \frac{i}{2}\Theta_\bk}
  \ket{\chi^\alpha_0}+ e^{\frac{i}{2}\Theta_\bk}e^{-i \delta\omega(k) t_0}
  \ket{\chi^\alpha_1}}{\sqrt{2}} , \no \end{eqnarray} where the two paths,
  $\nu=0,1$, may be subject to different dispersion with $\delta \omega(k)$
  being their difference.}\BibitemShut {Stop}%
\bibitem [{\citenamefont {Somma}\ \emph {et~al.}(2004)\citenamefont {Somma},
  \citenamefont {Ortiz}, \citenamefont {Barnum}, \citenamefont {Knill},\ and\
  \citenamefont {Viola}}]{somma-2004}%
  \BibitemOpen
  \bibfield  {author} {\bibinfo {author} {\bibfnamefont {R.}~\bibnamefont
  {Somma}}, \bibinfo {author} {\bibfnamefont {G.}~\bibnamefont {Ortiz}},
  \bibinfo {author} {\bibfnamefont {H.}~\bibnamefont {Barnum}}, \bibinfo
  {author} {\bibfnamefont {E.}~\bibnamefont {Knill}}, \ and\ \bibinfo {author}
  {\bibfnamefont {L.}~\bibnamefont {Viola}},\ }\href {\doibase
  10.1103/PhysRevA.70.042311} {\bibfield  {journal} {\bibinfo  {journal} {Phys.
  Rev. A}\ }\textbf {\bibinfo {volume} {70}},\ \bibinfo {pages} {042311}
  (\bibinfo {year} {2004})}\BibitemShut {NoStop}%
\bibitem [{\citenamefont {Ortiz}\ \emph {et~al.}(2005)\citenamefont {Ortiz},
  \citenamefont {Somma}, \citenamefont {Barnum}, \citenamefont {Knill},\ and\
  \citenamefont {Viola}}]{OrtizChapter}%
  \BibitemOpen
  \bibfield  {author} {\bibinfo {author} {\bibfnamefont {G.}~\bibnamefont
  {Ortiz}}, \bibinfo {author} {\bibfnamefont {R.}~\bibnamefont {Somma}},
  \bibinfo {author} {\bibfnamefont {H.}~\bibnamefont {Barnum}}, \bibinfo
  {author} {\bibfnamefont {E.}~\bibnamefont {Knill}}, \ and\ \bibinfo {author}
  {\bibfnamefont {L.}~\bibnamefont {Viola}},\ }in\ \href@noop {} {\emph
  {\bibinfo {booktitle} {Condensed Matter Theories}}},\ Vol.~\bibinfo {volume}
  {19},\ \bibinfo {editor} {edited by\ \bibinfo {editor} {\bibfnamefont
  {M.}~\bibnamefont {Belkacem}}\ and\ \bibinfo {editor} {\bibfnamefont {P.~M.}\
  \bibnamefont {Dinh}}}\ (\bibinfo  {publisher} {Nova Science Publishers,
  Inc.},\ \bibinfo {address} {Hauppauge, New York},\ \bibinfo {year} {2005})\
  pp.\ \bibinfo {pages} {297--308},\ \bibinfo {note}
  {quant-ph/0403043}\BibitemShut {NoStop}%
\bibitem [{\citenamefont {Gatti}\ \emph {et~al.}(2003)\citenamefont {Gatti},
  \citenamefont {Brambilla},\ and\ \citenamefont {Lugiato}}]{Gatti}%
  \BibitemOpen
  \bibfield  {author} {\bibinfo {author} {\bibfnamefont {A.}~\bibnamefont
  {Gatti}}, \bibinfo {author} {\bibfnamefont {E.}~\bibnamefont {Brambilla}}, \
  and\ \bibinfo {author} {\bibfnamefont {L.~A.}\ \bibnamefont {Lugiato}},\
  }\href {\doibase 10.1103/PhysRevLett.90.133603} {\bibfield  {journal}
  {\bibinfo  {journal} {Phys. Rev. Lett.}\ }\textbf {\bibinfo {volume} {90}},\
  \bibinfo {pages} {133603} (\bibinfo {year} {2003})}\BibitemShut {NoStop}%
\bibitem [{\citenamefont {Abouraddy}\ \emph {et~al.}(2004)\citenamefont
  {Abouraddy}, \citenamefont {Stone}, \citenamefont {Sergienko}, \citenamefont
  {Saleh},\ and\ \citenamefont {Teich}}]{Abouraddy}%
  \BibitemOpen
  \bibfield  {author} {\bibinfo {author} {\bibfnamefont {A.~F.}\ \bibnamefont
  {Abouraddy}}, \bibinfo {author} {\bibfnamefont {P.~R.}\ \bibnamefont
  {Stone}}, \bibinfo {author} {\bibfnamefont {A.~V.}\ \bibnamefont
  {Sergienko}}, \bibinfo {author} {\bibfnamefont {B.~E.~A.}\ \bibnamefont
  {Saleh}}, \ and\ \bibinfo {author} {\bibfnamefont {M.~C.}\ \bibnamefont
  {Teich}},\ }\href {\doibase 10.1103/PhysRevLett.93.213903} {\bibfield
  {journal} {\bibinfo  {journal} {Phys. Rev. Lett.}\ }\textbf {\bibinfo
  {volume} {93}},\ \bibinfo {pages} {213903} (\bibinfo {year}
  {2004})}\BibitemShut {NoStop}%
\bibitem [{\citenamefont {Ono}\ \emph {et~al.}(2013)\citenamefont {Ono},
  \citenamefont {Okamoto},\ and\ \citenamefont {Takeuchi}}]{Ono}%
  \BibitemOpen
  \bibfield  {author} {\bibinfo {author} {\bibfnamefont {T.}~\bibnamefont
  {Ono}}, \bibinfo {author} {\bibfnamefont {R.}~\bibnamefont {Okamoto}}, \ and\
  \bibinfo {author} {\bibfnamefont {S.}~\bibnamefont {Takeuchi}},\ }\href@noop
  {} {\bibfield  {journal} {\bibinfo  {journal} {Nat. Commun.}\ }\textbf
  {\bibinfo {volume} {4}},\ \bibinfo {pages} {2426} (\bibinfo {year}
  {2013})}\BibitemShut {NoStop}%
\bibitem [{\citenamefont {Gilaberte~Basset}\ \emph {et~al.}(2019)\citenamefont
  {Gilaberte~Basset}, \citenamefont {Setzpfandt}, \citenamefont {Steinlechner},
  \citenamefont {Beckert}, \citenamefont {Pertsch},\ and\ \citenamefont
  {Gr\"afe}}]{ReviewQI1}%
  \BibitemOpen
  \bibfield  {author} {\bibinfo {author} {\bibfnamefont {M.}~\bibnamefont
  {Gilaberte~Basset}}, \bibinfo {author} {\bibfnamefont {F.}~\bibnamefont
  {Setzpfandt}}, \bibinfo {author} {\bibfnamefont {F.}~\bibnamefont
  {Steinlechner}}, \bibinfo {author} {\bibfnamefont {E.}~\bibnamefont
  {Beckert}}, \bibinfo {author} {\bibfnamefont {T.}~\bibnamefont {Pertsch}}, \
  and\ \bibinfo {author} {\bibfnamefont {M.}~\bibnamefont {Gr\"afe}},\
  }\href@noop {} {\bibfield  {journal} {\bibinfo  {journal} {Laser \& Photonics
  Reviews}\ }\textbf {\bibinfo {volume} {13}},\ \bibinfo {pages} {1900097}
  (\bibinfo {year} {2019})}\BibitemShut {NoStop}%
\bibitem [{\citenamefont {Moreau}\ \emph {et~al.}(2019)\citenamefont {Moreau},
  \citenamefont {Toninelli}, \citenamefont {Gregory},\ and\ \citenamefont
  {Padgett}}]{ReviewQI2}%
  \BibitemOpen
  \bibfield  {author} {\bibinfo {author} {\bibfnamefont {P.~A.}\ \bibnamefont
  {Moreau}}, \bibinfo {author} {\bibfnamefont {E.}~\bibnamefont {Toninelli}},
  \bibinfo {author} {\bibfnamefont {T.}~\bibnamefont {Gregory}}, \ and\
  \bibinfo {author} {\bibfnamefont {M.~J.}\ \bibnamefont {Padgett}},\
  }\href@noop {} {\bibfield  {journal} {\bibinfo  {journal} {Nat. Rev. Phys.}\
  }\textbf {\bibinfo {volume} {1}},\ \bibinfo {pages} {367} (\bibinfo {year}
  {2019})}\BibitemShut {NoStop}%
\bibitem [{\citenamefont {Knill}\ \emph {et~al.}(2007)\citenamefont {Knill},
  \citenamefont {Ortiz},\ and\ \citenamefont {Somma}}]{KOS}%
  \BibitemOpen
  \bibfield  {author} {\bibinfo {author} {\bibfnamefont {E.}~\bibnamefont
  {Knill}}, \bibinfo {author} {\bibfnamefont {G.}~\bibnamefont {Ortiz}}, \ and\
  \bibinfo {author} {\bibfnamefont {R.~D.}\ \bibnamefont {Somma}},\ }\href
  {\doibase 10.1103/PhysRevA.75.012328} {\bibfield  {journal} {\bibinfo
  {journal} {Phys. Rev. A}\ }\textbf {\bibinfo {volume} {75}},\ \bibinfo
  {pages} {012328} (\bibinfo {year} {2007})}\BibitemShut {NoStop}%
\bibitem [{\citenamefont {de~Haan}\ \emph {et~al.}(2007)\citenamefont
  {de~Haan}, \citenamefont {Plomp}, \citenamefont {Bouwman}, \citenamefont
  {Trinker}, \citenamefont {Rekveldt}, \citenamefont {Duif}, \citenamefont
  {Jericha}, \citenamefont {Rauch},\ and\ \citenamefont {van
  Well}}]{deHaan2007}%
  \BibitemOpen
  \bibfield  {author} {\bibinfo {author} {\bibfnamefont {V.-O.}\ \bibnamefont
  {de~Haan}}, \bibinfo {author} {\bibfnamefont {J.}~\bibnamefont {Plomp}},
  \bibinfo {author} {\bibfnamefont {W.~G.}\ \bibnamefont {Bouwman}}, \bibinfo
  {author} {\bibfnamefont {M.}~\bibnamefont {Trinker}}, \bibinfo {author}
  {\bibfnamefont {M.~T.}\ \bibnamefont {Rekveldt}}, \bibinfo {author}
  {\bibfnamefont {C.~P.}\ \bibnamefont {Duif}}, \bibinfo {author}
  {\bibfnamefont {E.}~\bibnamefont {Jericha}}, \bibinfo {author} {\bibfnamefont
  {H.}~\bibnamefont {Rauch}}, \ and\ \bibinfo {author} {\bibfnamefont {A.~A.}\
  \bibnamefont {van Well}},\ }\href {\doibase 10.1107/S0021889806047558}
  {\bibfield  {journal} {\bibinfo  {journal} {Journal of Applied
  Crystallography}\ }\textbf {\bibinfo {volume} {40}},\ \bibinfo {pages} {151}
  (\bibinfo {year} {2007})}\BibitemShut {NoStop}%
\bibitem [{\citenamefont {Treimer}\ \emph {et~al.}(2006)\citenamefont
  {Treimer}, \citenamefont {Hilger},\ and\ \citenamefont
  {Strobl}}]{treimer2006}%
  \BibitemOpen
  \bibfield  {author} {\bibinfo {author} {\bibfnamefont {W.}~\bibnamefont
  {Treimer}}, \bibinfo {author} {\bibfnamefont {A.}~\bibnamefont {Hilger}}, \
  and\ \bibinfo {author} {\bibfnamefont {M.}~\bibnamefont {Strobl}},\
  }\href@noop {} {\bibfield  {journal} {\bibinfo  {journal} {Physica B:
  Condensed Matter}\ }\textbf {\bibinfo {volume} {385-386}},\ \bibinfo {pages}
  {1388} (\bibinfo {year} {2006})}\BibitemShut {NoStop}%
\bibitem [{\citenamefont {Majkrzak}\ \emph {et~al.}(2019)\citenamefont
  {Majkrzak}, \citenamefont {Berk}, \citenamefont {Maranville}, \citenamefont
  {Dura},\ and\ \citenamefont {Jach}}]{Majkrzak:2019yke}%
  \BibitemOpen
  \bibfield  {author} {\bibinfo {author} {\bibfnamefont {C.~F.}\ \bibnamefont
  {Majkrzak}}, \bibinfo {author} {\bibfnamefont {N.~F.}\ \bibnamefont {Berk}},
  \bibinfo {author} {\bibfnamefont {B.~B.}\ \bibnamefont {Maranville}},
  \bibinfo {author} {\bibfnamefont {J.~A.}\ \bibnamefont {Dura}}, \ and\
  \bibinfo {author} {\bibfnamefont {T.}~\bibnamefont {Jach}},\ }\href@noop {}
  {\  (\bibinfo {year} {2019})},\ \Eprint {http://arxiv.org/abs/1911.07974}
  {arXiv:1911.07974 [physics.ins-det]} \BibitemShut {NoStop}%
\end{thebibliography}%
\end{document}